\documentclass[final,3p,times,twocolumn]{elsarticle}

\usepackage{amssymb}
\usepackage{amsmath}
\usepackage{subcaption}
\usepackage{graphicx}
\usepackage{epstopdf}
\usepackage{xcolor}
\usepackage{physics}

\journal{International Journal of Non-Linear Mechanics}

\begin{document}

\begin{frontmatter}



\title{Vibrational resonance in vibro-impact oscillator through fast harmonic excitation}


\author[label1,label2]{Somnath Roy}
\affiliation[label1]{organization={Department of Applied Mechanics and Biomedical Engineering},
            addressline={Indian Institute of Technology Madras},
            city={Chennai},
            postcode={600036},
            state={Tamilnadu},
            country={India}}

\affiliation[label2]{organization={Centre for Complex Systems and Dynamics},
            addressline={Indian Institute of Technology Madras},
            city={Chennai},
            postcode={600036},
            state={Tamilnadu},
            country={India}}

\author[label1,label2]{Sayan Gupta} 

\begin{abstract}
This study focusses on extending the concept of  weak signal enhancement from dynamical systems based on vibrational resonance of nonlinear systems, to non-smooth systems. A Van der Pol- Duffing oscillator with a one sided barrier, subjected to harmonic excitations, has been considered as an archetypical low order model, whose response is weak. It is shown that that the system response can be significantly enhanced by applying an additional harmonic excitation but with much higher frequencies. The reasons for the underlying physics is investigated analytically using multiple-scale analysis and Blekham perturbation approach (direct partition motion). The analytical predictions are qualitatively validated using numerical simulations.  This approach yields valuable insights into the intricate interplay between fast and slow excitations in non-smooth systems.


\end{abstract}




\end{frontmatter}



\section{Introduction}


Resonance,  a well-studied phenomenon  in linear dynamical systems, occurs when the excitation frequency is equal or close to the natural frequency of the system, and leads to large amplitude oscillations. The increase in the oscillation amplitudes is due to phase synchronization of the forcing and system response, enabling efficient energy transfer from the forcing onto the system. Resonance is also observed in nonlinear systems but in a more complex manner.   The presence of geometric nonlinearities induces super-harmonics and sub-harmonics in the system response and can lead to 
super-harmonic and sub-harmonic resonance, internal resonance, and other modes of energy transfer \cite{nayfeh2008applied,manevich2005mechanics,vyas2009microresonator,das2023nonlinear,das2024effectsinternalresonancedamping}. 
These unique characteristics make nonlinear systems particularly interesting for studying resonance, which is a key objective in applications such as MEMS and piezoelectric resonators \cite{hajjaj2017mode,wang2019review}. In this context, a key mechanism for enhancing the system response to weak forcing in nonlinear dynamical systems involves the application of an additional fast harmonic signal drive. 
This approach referred to as vibrational resonance (VR), has been thoroughly explored over the past two decades \cite{yang2024vibrational}. VR has garnered significant interest in the scientific community and has evolved in various directions \cite{landa2000vibrational,gitterman2001bistable,blekhman2000vibrational,blekhman2004conjugate} 
due to its potential for widespread applications that are motivated either in enhancing the vibrations or in their mitigation.
This include studies 
in a multitude of systems, ranging from bistable oscillators \cite{baltanas2003experimental}, parametric oscillators \cite{roy2021vibrational,roy2023controlling}, time-delayed systems \cite{jeevarathinam2013effect}, nano-electromechanical systems \cite{chowdhury2020weak}, neuronal systems \cite{deng2010vibrational}, and many other diverse fields.

The fundamental mechanism of VR relies on three basic components:(i) geometric nonlinearity (ii) ``slow" excitations (characteristic signal), and (iii) ``fast" excitations (auxiliary signal). The terms ``fast" and ``slow" are qualitative and are used to compare the two excitation signals with respect to the time scales associated with the natural frequency of the system. Typically, the slowly varying excitation is generally considered to be weak. VR emerges through the interaction of fast forcing with the nonlinear components. In many instances, crucial information is embedded in a weak, low-frequency, or slowly varying signal, making the development of VR theory essential for detecting and amplifying these signals. The method of direct separation of motions \cite{blekhman2000vibrational} is commonly employed to analyze VR, typically to obtain responses at the excitation frequency of the slowly varying harmonic signal. To the best of the knowledge of the authors, while the aforementioned applications have been predominantly explored in smooth nonlinear systems, the application of vibrational resonance (VR) in impact systems, characterized by their inherently nonsmooth nature, presents a significant and largely untapped area for investigation.

A vibro-impact oscillator is characterized by its interaction with motion-limiting constraints, which can be either intermittent or continuous. This behavior is frequently encountered in diverse engineering applications across multiple fields. 
Examples of such systems include rotor-stator interactions in turbines, interacting gear teeth, pipes in heat exchangers,  riveted or bolted supports in structural systems \cite{ibrahim2009vibro,dimentberg2004random}, 
capsule systems used in endoscopy~\cite{gu2018dynamical}, drilling rigs employed in the oil and gas industry~\cite{de2019drill}, base isolation mechanisms in civil engineering~\cite{jangid2001base}, and critical infrastructure such as bridges~ \cite{dimitrakopoulos2013nonsmooth} and other strategic facilities \cite{kumar2015calculation}.
 The characteristic feature of these class of vibrating systems is the presence of partial restraints that allow free motion within domains confined by rigid barriers, but at the boundaries, the moving body impacts the barrier, leading to abrupt velocity reversals. This sudden change in velocity introduces nonsmooth nonlinearities [3,4] into the vibro-impact system, resulting in a rich phenomenological behavior across different parameter regimes. Vibro-impact systems exhibit a variety of dynamical behavior that is also observed in smooth systems, such as quasi-periodic oscillations \cite{luo2008periodic,zhang2017detecting,blazejczyk2001analysis}, Hopf bifurcations \cite{luo1998hopf,luo2002hopf,yue2008symmetry}, and chaotic dynamics \cite{de2008control,luo2001bifurcations,feng2015chaotic}, but also behavior that are specific to the non-smooth nature of these systems. The latter class includes phenomena such as border-collision bifurcation, grazing bifurcation, sliding bifurcation, and boundary-intersection crossing bifurcations. The effects of noise, typically referring to high frequency random signals that inevitably exist due to un-modelled characteristics, and which have the potential in significantly altering the dynamical behavior, have been investigated as well. This leads to additional bifurcation characteristics and which have been classified in the literature as stochastic bifurcations \cite{yang2017stochastic, kumar2016stochastic, kumar2017bifurcation, qian2021stochastic, rounak2020stochastic,makarenkov2012dynamics,belykh2023beyond}.

There exists a significant body of literature that have investigated the dynamical behavior of vibro-impacting systems to external harmonic excitations or in combination with random excitations, with the focus being on dynamics near strong resonance, super-harmonic resonance or in phenomena such as mode locking, phase locking, etc \cite{luo2007vibro,rong2010resonant,kember1999excitation,fritzkowski2021near,cho2013vibro,luo2008dynamics,nguyen2017new,haroun2015micro,haiwu2009subharmonic,superharmonic}. These studies show how 
pure harmonic driving forces or a mixture of harmonic and stochastic inputs influence the dynamic response, revealing complex behavior and potential resonance phenomena unique to vibro-impact systems. However, to the best of the authors' knowledge, there appear to be no studies on VR in vibro-impacting systems. 
 
Traditionally, vibrational resonance is achieved by tuning the external driving frequency close to the natural frequency of the oscillator. However, when the external forcing is too weak to induce a noticeable response, or when the forcing frequency is significantly different from the resonance frequency, or if experimental challenges arise due to the precision required in adjusting weak frequencies, it becomes essential to explore alternative methods for achieving resonance dynamics in vibro-impact oscillators.
For instance, in devices such as resonators, actuators, and certain MEMS devices, impacts are inherently part of their operation, making them functionally similar to vibro-impact oscillators \cite{tsai2022cmos}. The literature has explored these devices extensively, particularly regarding low-frequency energy harvesting and signal detection \cite{han2017study,liu2023harvesting,shi2015study,bo2014weak}. This presents an opportunity to apply VR as a robust method to enhance the response of systems to weak forcing. Such an approach could open new avenues for research in the domain of nonsmooth systems, offering an alternative strategy for enhancing system performance in these challenging regimes.
 
  This article is structured as follows: In the next section, the problem statement is presented. The VR phenomena in the vibro-impact oscillator is investigated analytically in Sec. 3. Section 4 presents a suite of numerical results and discussions that enable understanding of the behavior of the system, in the context of VR, and validates the analytical calculations qualitatively. The paper concludes in Sec. 5 where the salient features emerging from this study are presented.

\section{Problem Statement}

The Van der Pol-Duffing (VDPD) oscillator serves as a pivotal low order model for investigating dynamical systems with large geometric nonlinearities and with feedback, as is usually observed in any multi-physics interaction problems.
The VDPD oscillator exhibits a diverse array of dynamical behavior, including periodic, quasi-periodic, and chaotic oscillations 
\cite{venkatesan1997bifurcation,xu2020independent,wang2009response,kumar2017modified}, which can also be relevant to impacting systems. Its analytical and numerical tractability makes it a convenient and reliable model for theoretical studies and simulations.
VR in smooth VDPD oscillators have been recently investigated; see \cite{roy2021vibrational,roy2023controlling}. The focus of the present study is on investigating VR on non-smooth VDPD oscillators, with the non-smoothness being introduced by the presence of a one-sided barrier that leads to impacts. This has direct application in understanding the mechanism of detecting weak signals from various micromechanical devices \cite{chakraborty2021nonlinear}. 

\begin{figure}[h!]
    \centering
    \includegraphics[scale=0.8]{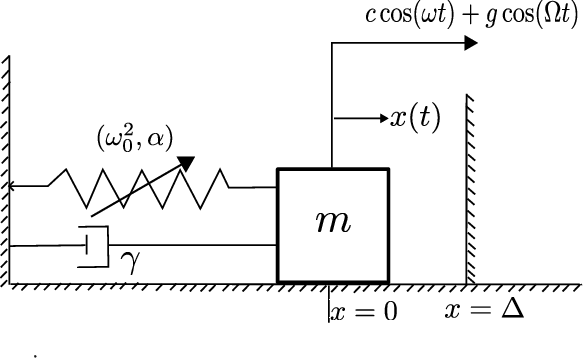}
    \caption{Schematic diagram of a vibro-impact oscillator, driven by bi-harmonic forces, with unilateral offset barrier at $x=\Delta$.}
    \label{fig0}
\end{figure}

The governing equation of a Van der Pol-Duffing (VDPD) oscillator, with a one-sided offset barrier at $x=\Delta$, driven by bi-harmonic forcing of slow and fast frequencies (Fig.\ref{fig0}), is given by
\begin{equation}
\ddot{x} + \gamma(x^2-1) \dot{x} - \omega_0^2 x + \alpha x^3 = c\cos \omega t+g \cos \Omega t, ~~x<\Delta,
\label{eq2.1}
\end{equation}
subjected to the impact condition
\begin{equation}
\Dot{x}^+=-e\Dot{x}^-,~~x=\Delta, ~~0< e\leq 1.
\label{eq2.2}
\end{equation}
Here \( x \), \( \dot{x} \), and \( \ddot{x} \) are respectively the 
displacement, velocity, and acceleration of the oscillator, 
\( \omega_0^2 \) is the coefficient of the linear stiffness (natural frequency), \( \alpha \) is the coefficient of the cubic term in the nonlinear 
stiffness of the system, and \( \gamma \) is the damping constant. On the right-hand side of the equation, there are two forcing terms \( c \cos(\omega t) \) and \( g \cos(\Omega t) \), where \( c \) and \( g \) are the amplitudes, and \( \omega \) and \( \Omega \) are the frequencies of the slow and fast forcing, respectively. To preserve the main essence of vibrational resonance (VR), it is assumed that \( \Omega \gg \omega, \omega_0 \). In Eq.~(\ref{eq2.2}), the impact condition is given at the one-sided barrier with a coefficient of restitution \( e \). The terms \( \dot{x}^+ \) and \( \dot{x}^- \) represent the tinstantaneous velocity of the system immediately after and immediately before the impact, respectively.

\section{Analytical formulation}
\subsection{Direct partition of motion}

To solve Eq.~(\ref{eq2.1}), the method of direct partition of motion is applied \cite{blekhman2000vibrational}. As evident from Eq.~(\ref{eq2.1}), the two forces acting on the system operate on widely different time scales: the slow scale is of the order $\mathcal{O}(\frac{1}{\omega})$, while the fast scale is of the order $\mathcal{O}(\frac{1}{\Omega})$. Therefore, 
the dynamical state variable 
$x(t)$ can be considered to be 
a superposition of a slow variable 
$X$ and a fast variable $\psi$, and can be expressed as 
\begin{equation}
    x(t)=X(t,\omega t)+\psi(t,\Omega t).
    \label{eq2.3}
\end{equation}
This mathematical artifice enables identifying the contributions of the fast and slow excitations and efficient book-keeping of their influence on the system response. The method of direct separation of motions involves two stages. Initially, as in the mechanics of systems with hidden motions, the original differential equations of the systems are transformed into a system of integro-differential equations of an order ``twice higher" with respect to the explicit (or ``slow'') components and hidden (or ``fast'') components, by selecting an appropriate expression for the additional (or `vibrational'') forces. In the second stage, the resulting system is solved 
by first approximating the fast forces as a $2\pi$-periodic function of the fast time scale, and subsequently averaging them out to obtain the effective slow motion of the system. The detailed rationale and mathematical abstraction behind this approach are available in\cite{blekhman2000vibrational}.

Substituting Eq.(\ref{eq2.3}) into Eq.(\ref{eq2.1}) and (\ref{eq2.2}) leads to rewriting the governing equation as 
\begin{equation}
\begin{split}
\ddot{X}&+\ddot{\psi}+\gamma(X^2+\psi^2+2X\psi-1)(\dot{X}+\dot{\psi})-\omega_{0}^{2}(X+\psi)\\
& +\alpha(X^3+3X^2\psi+3X\psi^2+\psi^3)=g\cos(\Omega t),
\end{split}
\label{eq2.4}
\end{equation}
with the constraint equation being recast as
\begin{equation}
    \Dot{X}^+=-e\Dot{X}^-,~~x=\Delta,
    \label{eq2.5}
\end{equation}
\begin{equation}
    \Dot{\psi}^+\approx\Dot{\psi}^-,~~x=\Delta.
    \label{eq2.6}
\end{equation}
In Eq. (\ref{eq2.6}), it is assumed that the fast variable 
$\psi$ and its derivatives ($\Ddot{\psi},\Dot{\psi}$) remain continuous during the impact. This assumption is justified by the fact that the frequency $\Omega$ is sufficiently high, rendering its time scale comparable to the impact time scale. It is now assumed that the fast varying part $\psi$ and its derivatives are all $2\pi$- periodic functions of the fast time scale $T_0=\Omega t$, with a zero mean. Mathematically, this implies that
\begin{equation}
\langle x(t)\rangle=\frac{1}{2\pi}\int_{0}^{2\pi}x\,~ \rm dT_0=X(t,\omega t),
\label{eq2.7}
\end{equation}
implying that on temporal averaging over time period $T_0$, (which is subsequently rescaled to $2\pi$), the smaller time scale components (fast) get averaged out, retaining only the large time scale component.  Here, $\langle \cdot \rangle$ denotes temporal average operator in time period $T_0$ (rescaled to $2\pi$).
Averaging Eq.(\ref{eq2.4}) yields:
\begin{equation}
\begin{split}
\ddot{X}&+\langle\ddot{\psi}\rangle+\gamma(X^2+\langle\psi^2\rangle+2s\langle\psi\rangle-1)(\dot{X}+\langle\dot{\psi}\rangle)\\
&-\omega_{0}^{2}(X+\langle\psi\rangle)+\alpha(X^3+3X^2\langle\psi\rangle\\
&+3X\langle\psi^2\rangle+\langle\psi^3\rangle)\\
&=c \cos (\omega t)+\langle g\cos(\Omega t)\rangle.
\label{eq2.8}
\end{split}
\end{equation}
Subtracting Eq.(\ref{eq2.8}) from Eq.(\ref{eq2.4}) and setting $\langle\ddot{\psi}\rangle=\langle\dot{\psi}\rangle=\langle\psi\rangle=0$, leads to the resulting equation 
\begin{equation}
\begin{split}
\ddot{\psi}&+\gamma((\psi^2-\langle\psi^2\rangle)\dot{X}+2X\psi\dot{X})\\
&+\gamma(X^2\dot{\psi}+\psi^2\dot{\psi}+2X\psi\dot{\psi}-\dot{\psi})\\
&-\omega_0^2\psi+\alpha(\psi^3-\langle\psi^3\rangle+3X^2\psi+3X(\psi^2-\langle\psi^2\rangle))\\
&=g\cos(\Omega t).
\label{eq2.9}
\end{split}
\end{equation}
%
As the second derivative of the fast variable dominates the first derivative and the state variable  $(i.e.,\ddot{\psi}>>\dot{\psi}>\psi,\psi^2,\psi^3...)$, this is referred to as the \textit{inertial approximation} \cite{blekhman2000vibrational}. The governing equation can now be approximated  to the simpler form  as $\ddot{\psi}\approx g\cos(\Omega t)$, which in turn, leads to approximating the fast response as
$\psi\approx-\frac{g}{\Omega^2}\cos(\Omega t)$. This leads to the following statistical properties: $\langle\psi^2\rangle=\frac{g^2}{2\Omega^4}$ and $\langle\psi\rangle=\langle\psi^3\rangle=0$ Consequently, the equation governing the slow motion is derived as
\begin{equation}
\ddot{X}+\gamma(X^2-K)\dot{X}+\tilde{\omega}^2X+\alpha X^3=c\cos(\omega t).
\label{eq2.10}
\end{equation}
Eq. (\ref{eq2.10}) describes the \textit{effective slow dynamics} of the original system described by Eq.(\ref{eq2.1}). It is important to note that the influence of the amplitude of the fast forcing is in modifying the linear stiffness by coupling with the nonlinear stiffness, thus appearing as an effective natural frequency, given by
\begin{equation}
    \Tilde{\omega}=\sqrt{\frac{3\alpha g^2}{2\Omega^4}-\omega_0^2}
    \label{eq2.11}
\end{equation}
It is seen that the effective frequency is a function of the fast forcing strength 
$g$ and frequency $\Omega$. This expression also helps in appreciating that by tuning $g$, the effective frequency can be brought closer to $\omega$, thereby achieving the maximum response in the system. 
It is important to note that the nonlinearity in the damping term is also modified by the fast forcing,  the relationship  of which is given by 
\begin{equation}
    K=1-\frac{g^2}{2\Omega^4}.
    \label{eq2.12}
\end{equation}
In the next part, an analytical description of the response amplitude is presented. By employing the multiple-scale perturbation method \cite{nayfeh2008applied}, the flow equation is derived, which enables expressing the amplitude as a function of the fast forcing strength.

\subsection{Formulation of response amplitude}
To proceed further, the multiple-scale perturbation method is applied to Eq. (\ref{eq2.10}) along with the boundary condition in   Eq.(\ref{eq2.5}).  Introducing a perturbation parameter $\epsilon$, such that $\gamma=\epsilon \delta, \alpha=\epsilon\lambda, c=\epsilon f$ and  dimensionless time $\omega t=\tau$, Eq.(\ref{eq2.10}) can be rewritten as  
\begin{equation}
X''+\epsilon\frac{\delta}{\omega}(X^2-K)X'+\frac{\tilde{\omega}^2}{\omega^2}X+\epsilon\frac{\lambda}{\omega^2}X^3=\epsilon f\cos\tau.
\label{eq2.13}
\end{equation}
Here, the $'$ denotes the derivative with respect to  
$\tau$. 
Primary resonance is investigated by setting $\omega=\Tilde{\omega}+\epsilon\Tilde{\sigma}$, where $\Tilde{\sigma}$ represents the detuning parameter. The following new set of parameters are defined: $\Gamma=\frac{\delta}\omega, \Lambda=\frac{\lambda}{\omega^2}$, $F=\frac{f}{\omega^2}$  and  
\begin{equation}
\frac{\tilde{\omega}^2}{\omega^2}=1-\epsilon\sigma,
 \label{eq2.14}
\end{equation}
where $\sigma=\frac{2\Tilde{\sigma}\Tilde{\omega}}{\omega^2}.$
Eq.(\ref{eq2.13}) 
can now be rearranged as
\begin{equation}
   X''+\epsilon\Gamma(X^2-K)X'+X+\epsilon\Lambda X^3=\epsilon F\cos\tau+\epsilon\sigma X.
   \label{eq2.15}
\end{equation}
The application of multiple time scales begins with  the expansion of variable $X$ in powers of $\epsilon$ as
\begin{equation}
    X(\tau_0,\tau_1)=X_0(\tau_0,\tau_1)+\epsilon X_1(\tau_0,\tau_1)+\mathcal{O}(\epsilon^2) \hdots
    \label{eq2.16}
\end{equation}
Here, two distinct time scales are considered: a fast time scale $\tau_0$ and a slow time scale $\tau_1$. More generally, these time scales and their corresponding derivatives are defined as follows:
\begin{equation}
\begin{split}
\tau_n&=\epsilon^n\tau;\\
\frac{\rm d}{\rm d\tau}&=\frac{\partial}{\partial\tau_0}+\epsilon\frac{\partial}{\partial\tau_1}\equiv D_0+\epsilon D_1,\\
\frac{\rm d^2}{\rm d\tau^2}&=\frac{\partial^2}{\partial\tau_0^2}+2\epsilon\frac{\partial^2}{\partial\tau_1 \partial\tau_0}+\epsilon^2\frac{\partial^2}{\partial\tau_1^2}\\
&\equiv D_0+2\epsilon D_0D_1+\epsilon^2D_2.
\end{split}
\label{eq2.17}
\end{equation}
Substituting Eq.(\ref{eq2.16}) into Eq.(\ref{eq2.15}) and by using Eq.(\ref{eq2.17}), the equations can be decomposed according to the order of $\epsilon$. Consequently, the zeroth-order $\mathcal{O}(\epsilon^0)$ equation can be expressed as
\begin{equation}
    X_0''+X_0=0,
    \label{eq2.18}
\end{equation}
the solution for which is given by
\begin{equation}
X_0(\tau_0,\tau_1)=C_1(\tau_1)\cos\tau_0+C_2(\tau_1)\sin\tau_0.
    \label{eq2.19}
\end{equation}
Here, $C_1$ and $C_2$ are arbitrary constants, presumed to be the function of slow time $\tau_1$, and are determined from 
the boundary conditions, particularly the impact conditions. Starting from the instant immediately following impact,  the boundary conditions for all periodic motion combinations can be expressed as
$$X_0(0,0)=\Delta,$$ 
$${X_0'}={X_0'}^+$$  and  $$X_0(\frac{2\pi l}{\omega},\frac{2\pi l}{\omega})=\Delta,$$  $${X_0'}={X_0'}^-,$$ where $l$ is regime multiple. Substituting these conditions in Eq. (\ref{eq2.19}) results in 
\begin{equation}
C_1(0)=C_1\Bigg(\frac{2\pi l}{\omega}\Bigg)=\Delta ~~\textrm{and} ~~ C_2(0)=C_2\Bigg(\frac{2\pi l}{\omega}\Bigg)=\Delta\tan\Bigg(\frac{\pi l}{\omega}\Bigg).
\label{eq2.20}
\end{equation}
Eq.(\ref{eq2.19}) can be written in polar form as
\begin{equation}
X_0(\tau_0,\tau_1)=R(\tau_1)\cos(\tau_0+\theta(\tau_1)),
    \label{eq2.21}
\end{equation}
where
\begin{eqnarray}
R(\tau_1) &= &\sqrt{C_1^2(\tau_1)+C_2^2(\tau_1)},\\
\nonumber
\theta(\tau_1) & =& \tan^{-1}\Bigg(-\frac{C_2}{C_1}\Bigg),\\
 \label{eq2.22}
\end{eqnarray}
from which the initial amplitude and phase are defined as 
$$R(0)=R_0=\Delta\sec\bigg(\frac{\pi l}{\omega}\bigg)$$ and $$\theta(0)=\theta_0=\tan^{-1}\Bigg(-\tan\frac{\pi l}{\omega}\Bigg).$$
The first-order $\mathcal{O}(\epsilon^1)$ equation can now be expressed as
\begin{equation}
\begin{split}
X_1''&+X_1=-2D_1D_0X_0-\Gamma(X_0^2-K)D_0X_0-\Lambda X_0^3\\
&+F\cos\tau_0+\sigma X_0\\
&=\bigg[2D_1R\sin(\tau_0+\theta)+2RD_1\theta\cos(\tau_0+\theta)\bigg]\\
&+\frac{1}{4}\Gamma R^3\sin(\tau_0+\theta)+\frac{1}{4}\Gamma R^3\sin(3\tau_0+3\theta)\\
&-\Gamma KR\sin(\tau_0+\theta)-\frac{3}{4}\Lambda R^3\cos(\tau_0+\theta)\\
&-\frac{1}{4}\Lambda R^3\cos(3\tau_0+3\theta)+F\cos(\tau_0+\theta)\cos\theta\\
&+F\sin(\tau_0+\theta)\sin\theta+\sigma R\cos(\tau_0+\theta).
\end{split}
 \label{eq2.23}
\end{equation}
When solving for $X_1$, the terms involving 
$\sin(\tau_0+\theta)$ and $\cos(\tau_0+\theta)$ yield a divergent solution, contradicting the physical situation. To eliminate these secular terms, the coefficients of $\sin(\tau_0+\theta)$ and $\cos(\tau_0+\theta)$ must be set to zero, thereby providing the desired slow flow equations as
\begin{equation}
\begin{split}
\frac{\partial R}{\partial \tau_1}&=-\frac{\Gamma R^3}{8}+\frac{\Gamma K R}{2}-\frac{F}{2}\cos\theta,\\
R\frac{\partial\theta}{\partial \tau_1}&=\frac{3\Lambda R^3}{8}-\frac{\sigma}{2} R-\frac{F}{2}\sin\theta.
\end{split}
 \label{eq2.24}
\end{equation}
Now Eq.(\ref{eq2.23}) is simplified by discarding the diverging terms,  leading to the following simplified form as
\begin{equation}
X_1''+X_1= \frac{1}{4}\Gamma R^3\sin(3\tau_0+3\theta)-\frac{1}{4}\Lambda R^3\cos(3\tau_0+3\theta).
 \label{eq2.25}
\end{equation}
The corresponding general solution can now be expressed as
\begin{equation}
\begin{split}
X_1(\tau_0,\tau_1)&=R_1(\tau_1)\cos(\tau_0+\theta_1(\tau_1))\\
&-\frac{1}{32}\Gamma R^3\sin(3\tau_0+3\theta)\\
&+\frac{1}{32}\Lambda R^3\cos(3\tau_0+3\theta).
\end{split}
 \label{eq2.26}
\end{equation}
To determine the dependency of $R_1$ and $\theta_1$ on the slow time scale, a second-order calculation is required to formulate the flow equations for $R_1$ and $\theta_1$. 
This level of detail is beyond the scope of this article, but the analytical expression for the amplitude can be derived from Eq.(\ref{eq2.24}) by considering the stable oscillation of the system. For a stable fixed point $(R^*,\theta^*)$, the condition $\frac{\partial R}{\partial\tau_1}=\frac{\partial\theta}{\partial\tau_1}=0$ is imposed, and using trigonometric identity, Eq.(\ref{eq2.24}) gives the solution of $\sigma$ in the form
\begin{equation}
\sigma=\frac{3}{4}\Lambda R^2\pm\sqrt{\frac{\Gamma^2 R^2}{2}+\frac{F^2}{R^2}-\frac{\Gamma^2R^4}{16}-\Gamma^2K^2}.
\label{eq2.27}
\end{equation}
Substituting the value of $\sigma$ into  Eq.(\ref{eq2.14}), an analytical expression,  
 elucidating the relationship between the response amplitude $R$ and the strength of fast excitation $g$, given by
\begin{equation}
g=\Omega^2\left[\frac{2}{3\alpha}[\omega_0^2+\omega^2\left(1-\epsilon\sigma\right)]\right]
\label{eq2.28}
\end{equation}
is derived. This relation shows 
 the variation of the amplitude $R$ 
 with 
 $g$, providing a clear analytical insight into the 
 the behavior of the system under fast excitation. 
In the subsequent section, an in-depth discussion of the numerical results is provided, followed by a comparison with the  analytical predictions.

\section{Numerical Results}
\noindent To validate the analytical results, numerical simulations are conducted on the original system incorporating the impact condition (Eqs.\eqref{eq2.1} and \eqref{eq2.2}). The original numerical approach of VR as described in \cite{landa2000vibrational} was carried out by assuming a smooth Duffing oscillator. The primary objective is to extract the Fourier 
components 
from  $x(t)$,  
the response of the nonlinear system which comprises multiple harmonics. The response $x(t)$ has been simulated for a long time and the time series analysis is carried out by discarding the initial segment of the time history, to ensure steady-state conditions have been reached.

\begin{figure}[h]
\centering
\includegraphics[scale=0.7]{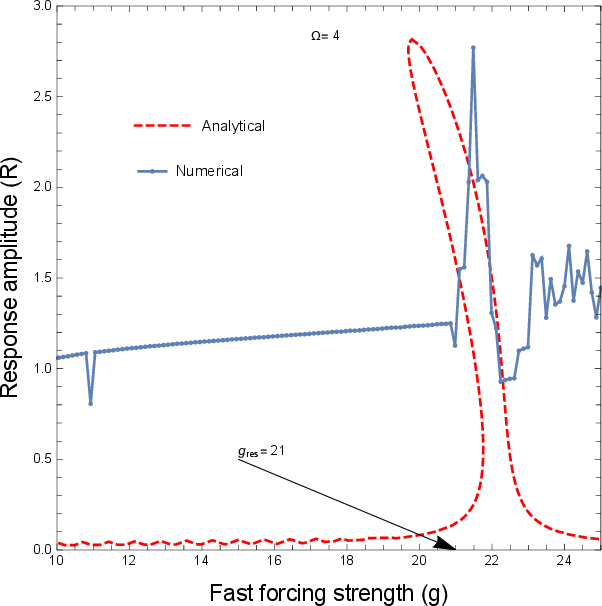}
\caption{Resonance response $R$ of the system is shown by varying the fast forcing amplitude $g$ at fast forcing frequency $\Omega=4$. The impact barrier is at $\Delta=1.0$ and restitution coefficient $e=0.95$.Other parameters are taken as:$\gamma=0.22,\alpha=0.1,c=0.04,\omega_0=0.1,\omega=0.5$. Numerical simulation (dotted line) is based on (Eqs.\eqref{eq2.1} and \eqref{eq2.2}) where the analytical curve (dashed line) is derived from Eq.(\ref{eq2.28}). Both analytical and numerical results are in good agreement, consistently indicating a resonance peak at $g_{res}=21$ as predicted by Eq.(\ref{eq3.4})}
\label{fig1}
\end{figure}

\begin{figure}[t]
\centering
\includegraphics[scale=0.7]{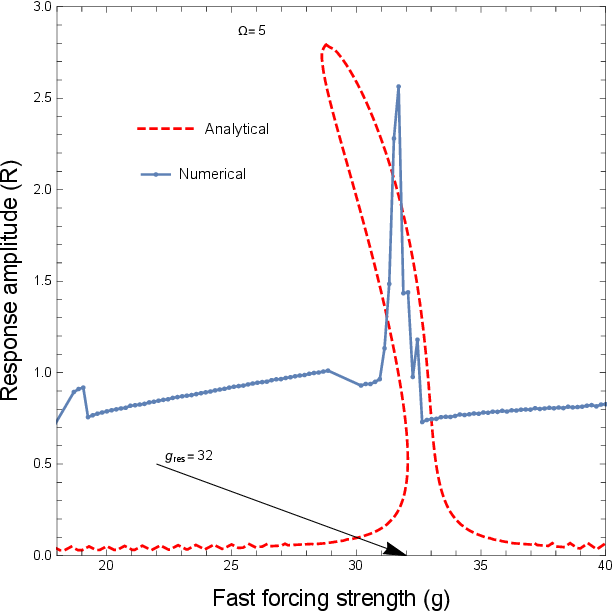}
\caption{Resonance response $R$ of the system is shown by varying the fast forcing amplitude $g$ at fast forcing frequency $\Omega=5$. Impact barrier is at $\Delta=1.0$ and restitution coefficient $e=0.95$.Other parameters are taken as:$\gamma=0.22,\alpha=0.1,c=0.04,\omega_0=0.1,\omega=0.5$. Numerical simulation (dotted line) is based on (Eqs.\eqref{eq2.1} and \eqref{eq2.2}) where the analytical curve (dashed line) is derived from Eq.(\ref{eq2.28}). Both analytical and numerical results are in good agreement, consistently indicating a resonance peak at $g_{res}=32$ as predicted by Eq.(\ref{eq3.4})}
\label{fig2}
\end{figure}

The adaptive Runge-Kutta 4th order (RK-4) method is utilized for numerical integration. When an impact occurs, the integration is halted, and the velocity is reversed according to Eq.(\ref{eq2.2}). The updated velocity serves as the new initial condition for the subsequent integration. To enhance accuracy near the impact, adaptive time steps have been taken. Consequently, the sine and cosine Fourier components can be expressed as
\begin{eqnarray}
 Q_s(\omega)=\frac{2}{mT}\int_{0}^{mT} x(t)\sin(\omega t){\rm d}t,
 \label{eq3.1}\\
 Q_c(\omega)=\frac{2}{mT}\int_{0}^{mT} x(t)\cos(\omega t){\rm d}t,
 \label{eq3.2}
\end{eqnarray}
where $m$ is a sufficiently large integer. The response amplitude is 
numerically computed  as 
\begin{eqnarray}
R(\omega)=\frac{\sqrt{Q_{s}^2(\omega)+Q_{c}^2(\omega)}}{c}
\label{eq3.3}
\end{eqnarray}

The numerically computed response is plotted against the analytical result Eq.(\ref{eq2.28}), as illustrated in Fig.\ref{fig1} and Fig.\ref{fig2}, where the phenomenon of resonance has been demonstrated in this type of vibro-impact oscillator by tuning the fast forcing strength $g$. This is validated by the analytically predicted value of $g$, specifically $g_{res}$, at which the resonance peak is observed. Also, both the analytical and numerical approaches indicate nearly identical maximum amplitudes. The value of $g_{res}$ can be obtained from Eq.(\ref{eq2.11}) when the effective frequency $\tilde{\omega}$ closely matches the slow drive frequency $\omega$, as
\begin{equation}
g_{res}=\Omega^2\sqrt{\frac{2}{3\alpha}(\omega^2+\omega_0^2)}
\label{eq3.4}
\end{equation}
The numerical values of the parameters are taken to be
$\gamma=0.22,\alpha=0.1,c=0.04,\omega_0=0.1,\omega=0.5, \Delta=1.0$ and $e=0.95$. For $\Omega=4$, it is observed that $g_{res}=21$ (see Fig.\ref{fig1}) and when $\Omega=5$, the corresponding value is seen to be $g_{res}=32$ in Fig.\ref{fig2}.
Notably, as the fast frequency $\Omega$ increases, the resonance peak shifts to higher values of $g$, as expected. 
Even though the numerical results appear to be at wide variance with the analytical results over much of the range, the values of $g$ at which higher response is observed analytically is validated through numerical simulations.  Additionally, the amplitude at VR are observed to be of the same order.

An important aspect of analytical solutions needs to be addressed, irrespective of the numerical results at hand. The practical advantages of analytical solutions are particularly significant in experimental settings. Knowing in advance where to expect the maximum amplitude can save considerable time when detecting weak signals. Furthermore, analytical expressions explicitly reveal the interplay between system parameters, which is crucial for real-life applications. For example, the analytical results show that the fast forcing strength at which the maximum peak appears $g_{res}$ is proportional to the fast frequency $\Omega$ but inversely proportional to the coefficient of nonlinear stiffness $\alpha$. While numerical procedures yield robust results, the insights provided by analytical solutions are invaluable for understanding and optimizing system behavior.

\begin{figure}[h]
    \centering
    \subcaptionbox{}{\hspace{-1cm}\includegraphics[height=5.8cm,width=10.3cm]{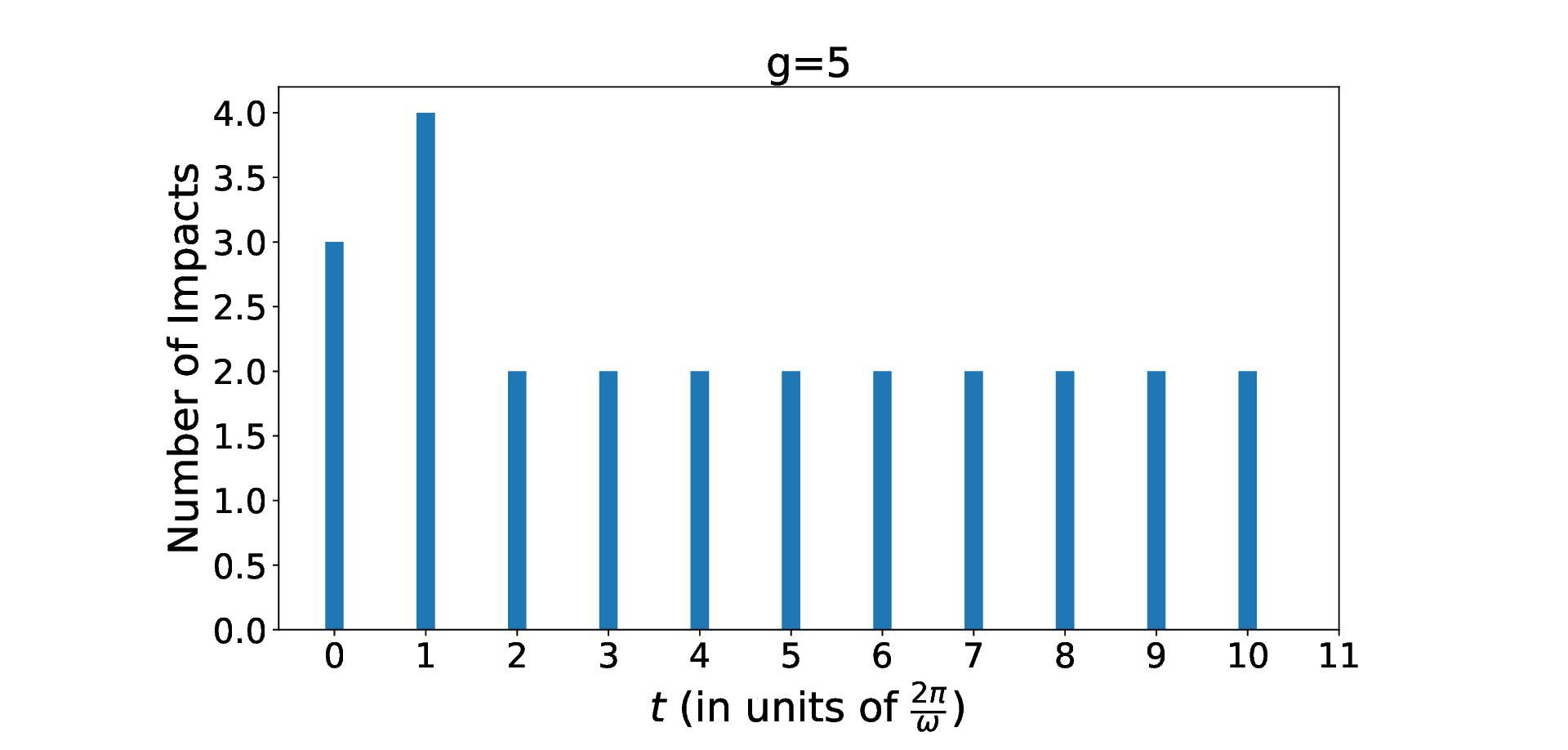}}
    \subcaptionbox{}{\hspace{-1cm}\includegraphics[height=5.8cm,width=10.3cm]{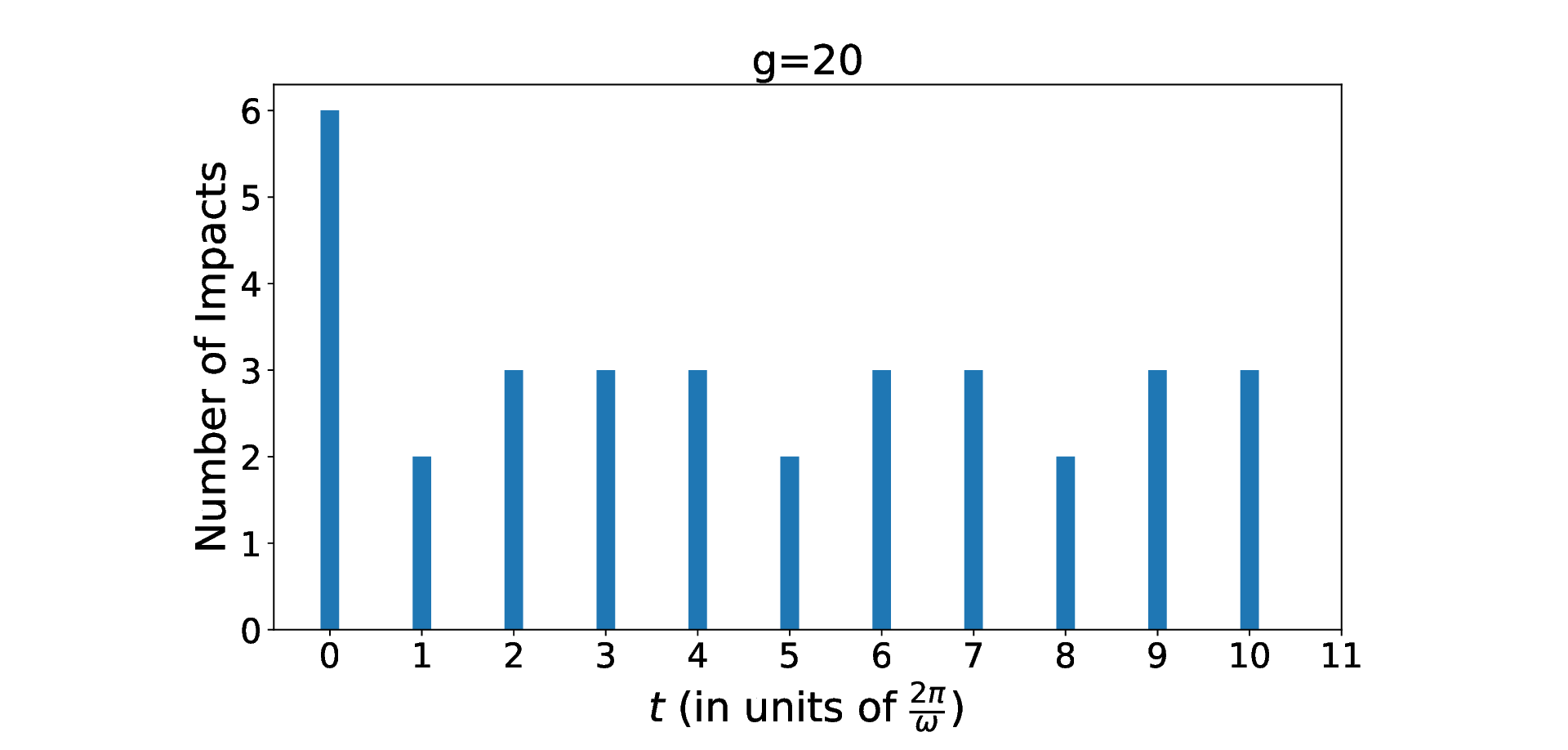}}
    \caption{Number of impacts is plotted against the period of $\frac{2\pi}{\omega}$ simulated from the original system (Eqs.\eqref{eq2.1} and \eqref{eq2.2}) . The fast forcing strength (a) for $g=5$ (b) for $g=20$. The fast frequency $\Omega=5$ and the slow frequency $\omega=0.5$ are taken for both scenarios. }
    \label{impact_no}
\end{figure}
Although the analytical expression in Eq.(\ref{eq2.28}) shows reasonable agreement with the numerical results in Fig.\ref{fig1} and Fig.\ref{fig2}, discrepancies arise at higher values of $g$ due to three primary reasons.
First, the analytical approach involves decomposing the system dynamics into fast and slow time scales, averaging out the rapidly varying components to obtain effective slow dynamics. In contrast, numerical integration retains all time scales, capturing the complete dynamics.
Secondly, in deriving the flow equations Eq.(\ref{eq2.24}), the calculations are restricted to terms of order $\mathcal{O}(\epsilon)$. The numerical method, however, can capture the behavior of the system 
with higher accuracy, effectively accounting for higher-order effects. This inherent limitation is a common challenge for all perturbative approaches.
Finally, the 
inherent nonsmoothness of the system introduces 
further complexities. To attain resonance, it is necessary to increase the fast forcing strength $g$. This increase leads to multiple impacts within a period $\frac{2\pi}{\omega}$. The effect of multiple impacts can be analyzed by simulating the original dynamics of the system described by Eqs.\eqref{eq2.1} and \eqref{eq2.2}, as illustrated in Fig.\ref{impact_no}.
This figure plots the number of impacts against periods of $\frac{2\pi}{\omega}$
corresponding to two distinct cases of fast forcing strength; when $g=5$ and when 
$g=20$, respectively. An inspection of this figure reveals that irrespective of the the value of $g$, the number of impacts per period is higher initially but stabilizes with large time. However, it is evident that for  higher values of $g$, the number of impacts is higher not only during transience but also when the system has stabilized.
%
This phenomenon also explains why the initial amplitude is higher in the numerical results compared to the analytical plots in Figs. \ref{fig1} and \ref{fig2}. In the analytical approach, the fast frequency components are averaged out, leading to a discrepancy between the initial amplitudes observed in the numerical and analytical results. Additionally, the evolving dynamics at the impact boundary are intricate, necessitating an optimized numerical scheme to achieve reasonable results. During the initial stage of the numerical simulations, transient fluctuations are observed, which should be removed to accurately capture the 
\begin{figure*}
\centering
    \subcaptionbox{}{\includegraphics[scale=0.16]{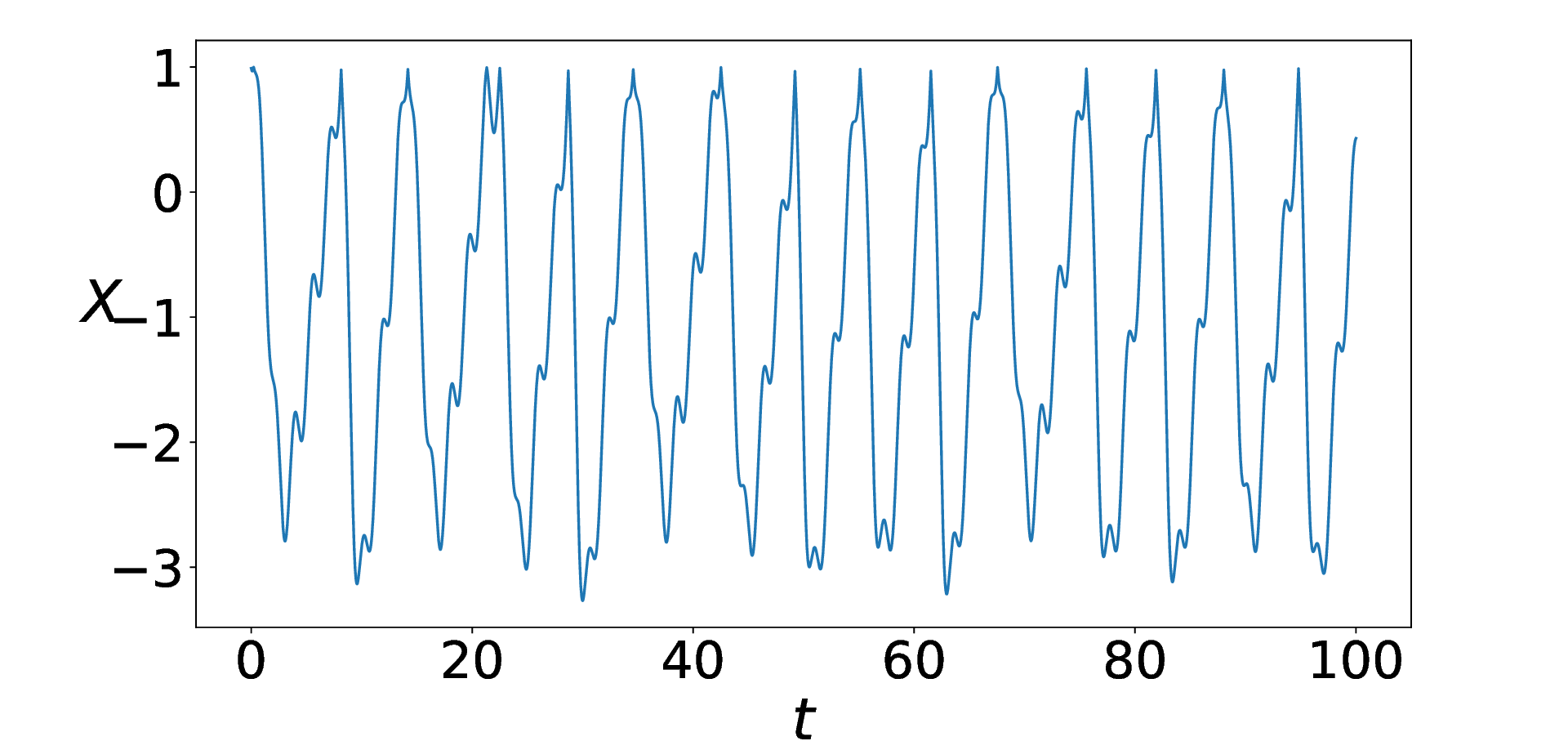}}
    \subcaptionbox{}{\includegraphics[scale=0.16]{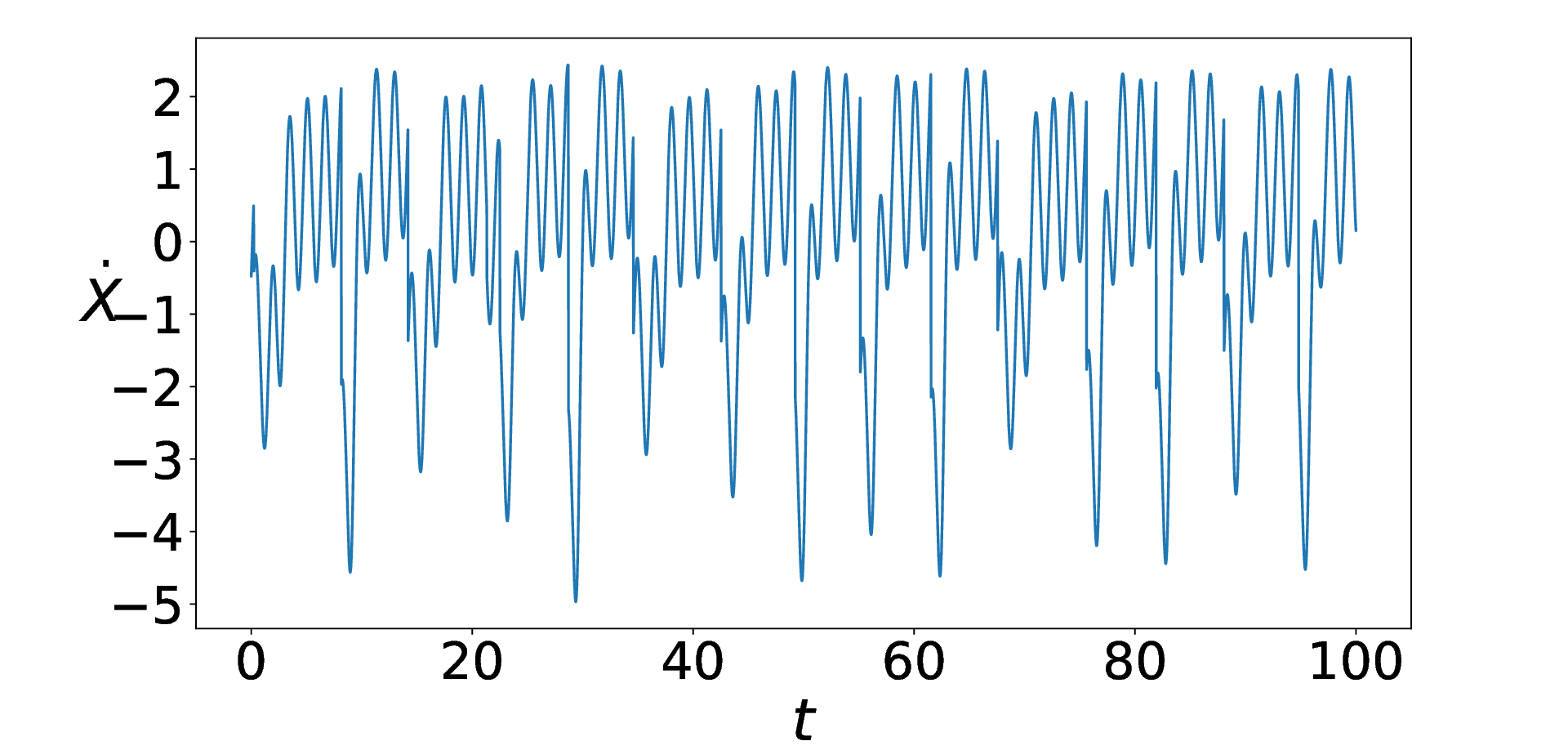}}
    \subcaptionbox{}{\includegraphics[scale=0.16]{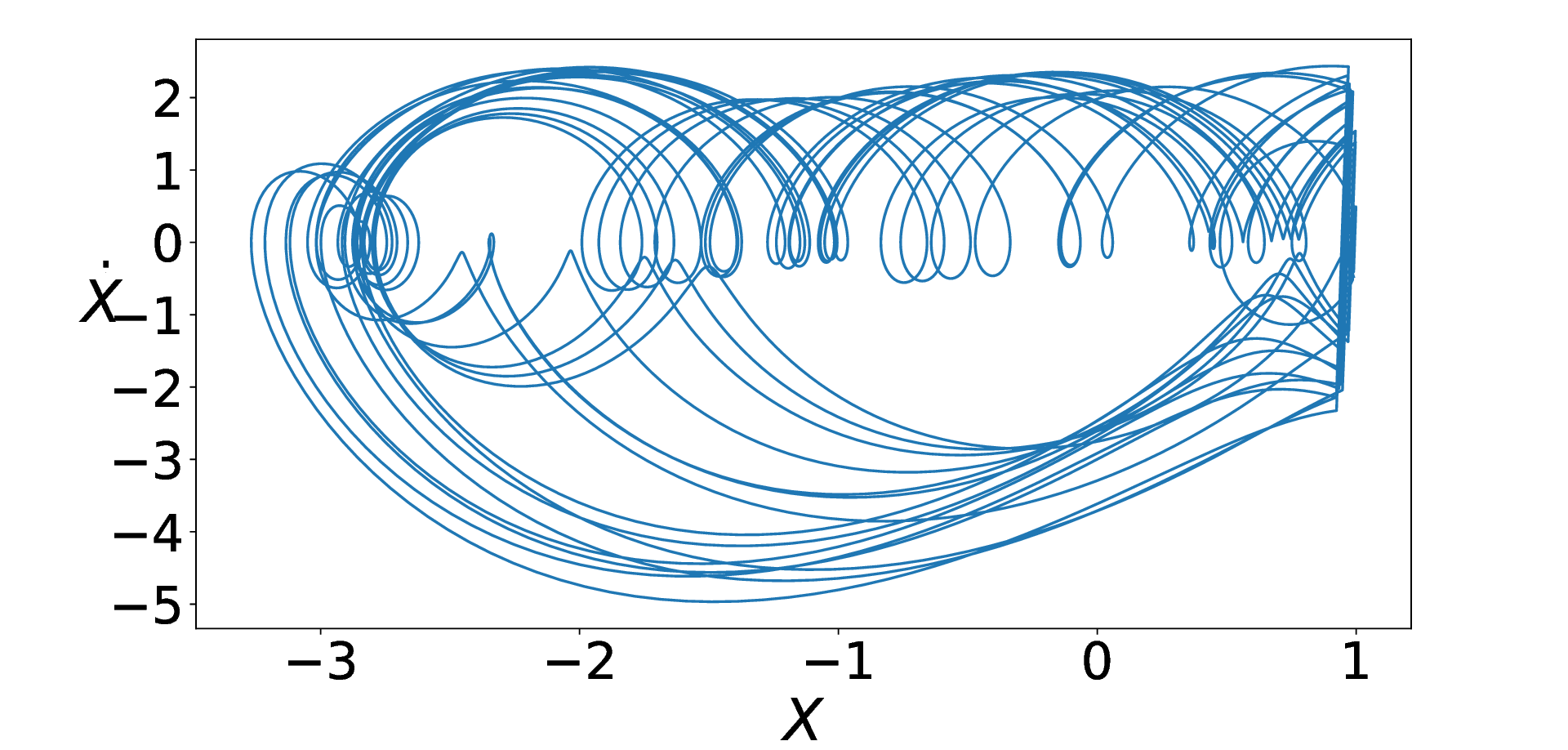}}

   \subcaptionbox{}{\includegraphics[scale=0.16]{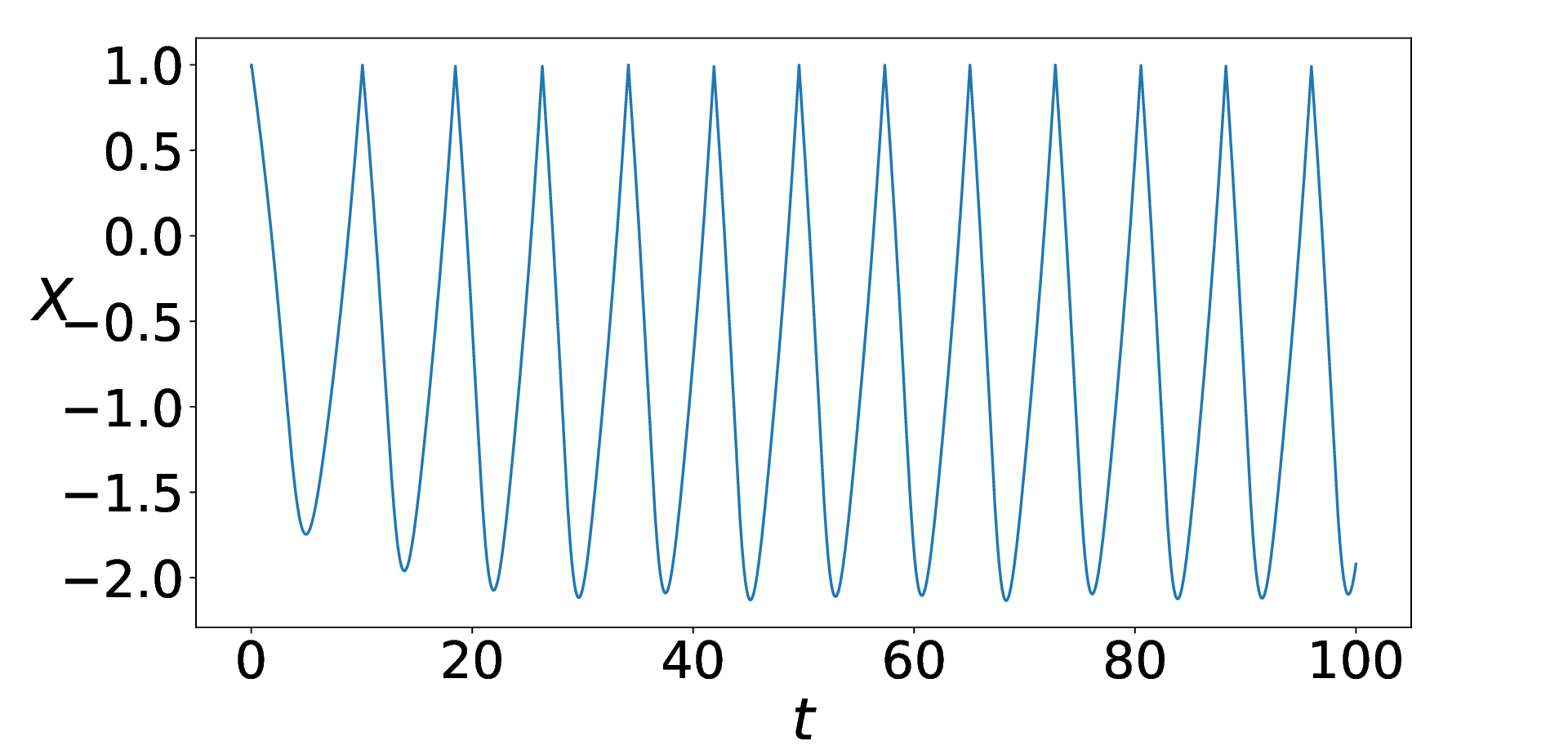}}
    \subcaptionbox{}{\includegraphics[scale=0.16]{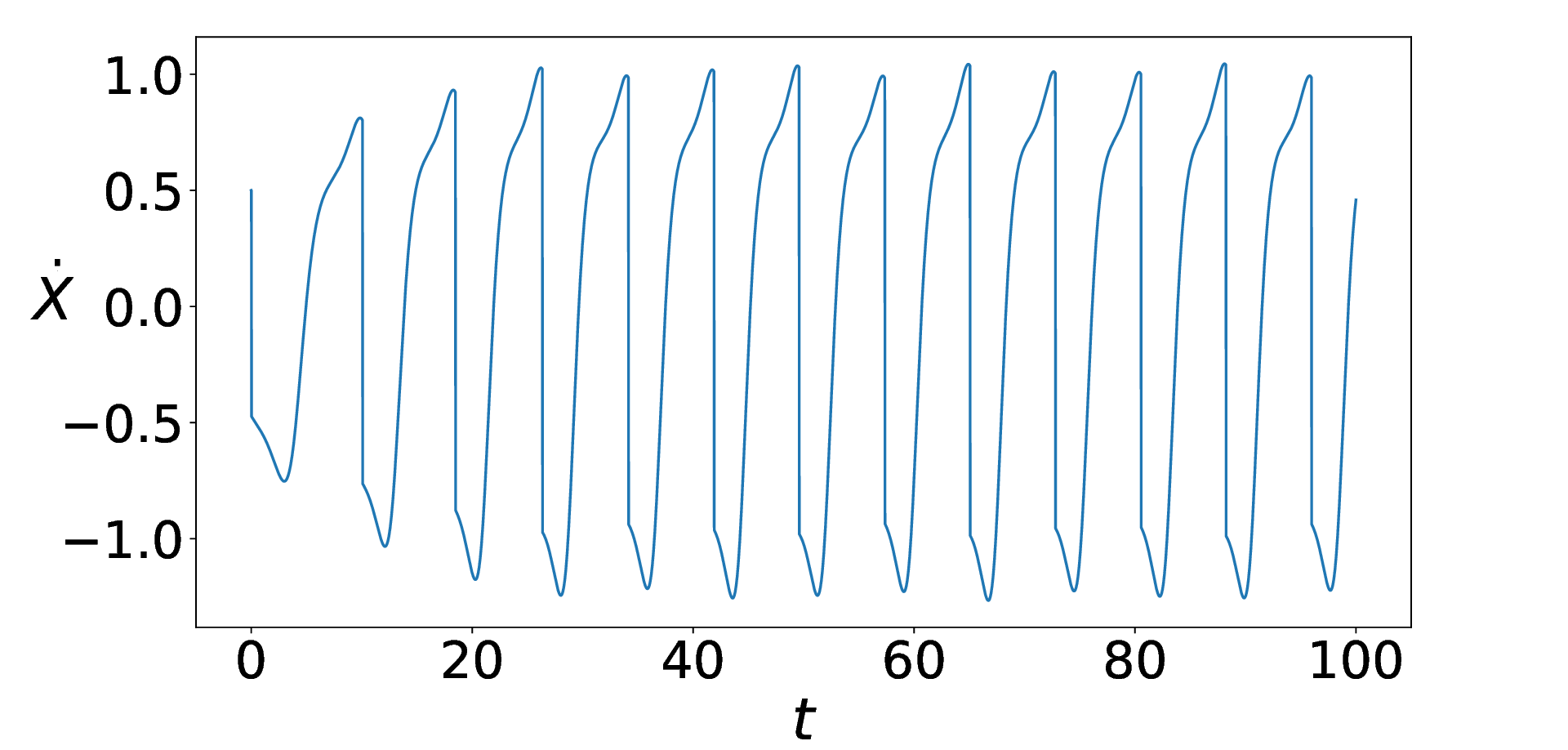}}
    \subcaptionbox{}{\includegraphics[scale=0.16]{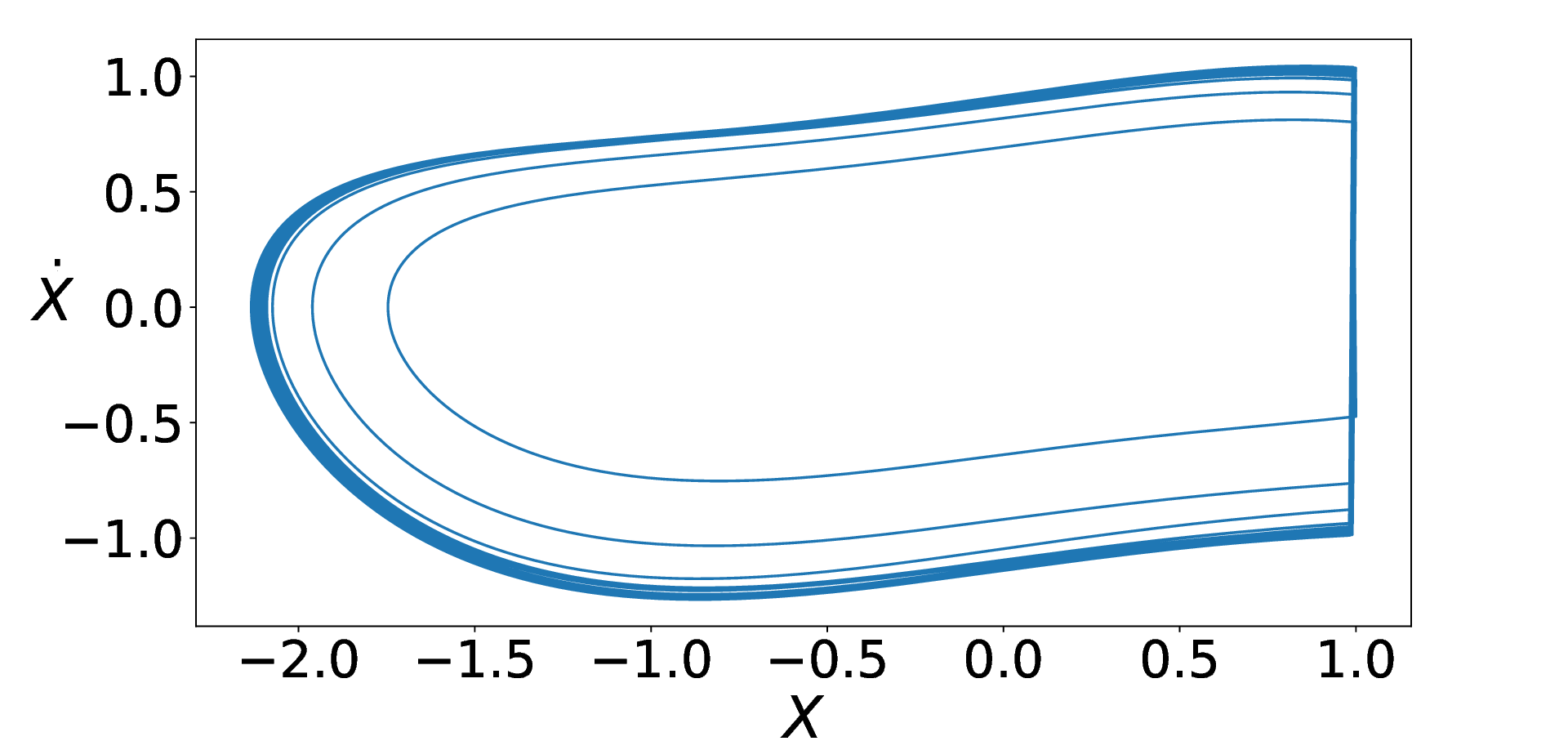}}
\caption{Comparison of original dynamics and effective dynamics for $g=5$ and $\Omega=4$. First row: simulations are based on the original system (Eqs.\eqref{eq2.1} and \eqref{eq2.2}). Second row: simulations are based on the effective system Eq.(\ref{eq2.10}). The pairs (a,d) show the position history, (b,e) display the velocity history, and (c,f) present the phase plot of the corresponding main and effective slow dynamics.The other parameters are set at: $\gamma=0.22,\alpha=0.1,c=0.04,\omega_0=0.1,\omega=0.5, \Delta=1.0$ and $e=0.95.$ The initial values of position and velocity are chosen as $x(0)=0.99$ and $\dot{x}(0)=0.5$.  }
\label{fig3}
\end{figure*}
steady-state behavior of the system. In summary, conflicts arise from the separation of the time scale and averaging out the fast scale. This is followed by neglecting higher-order corrections in perturbative calculations, and finally, it involves the multiple impacts and chattering induced by the rapid frequency which can be realized by observing the original dynamics of the system and shown in Figs.\ref{fig3}, \ref{fig4} and \ref{fig5}.

\begin{figure*}
\centering
    \subcaptionbox{}{\includegraphics[scale=0.16]{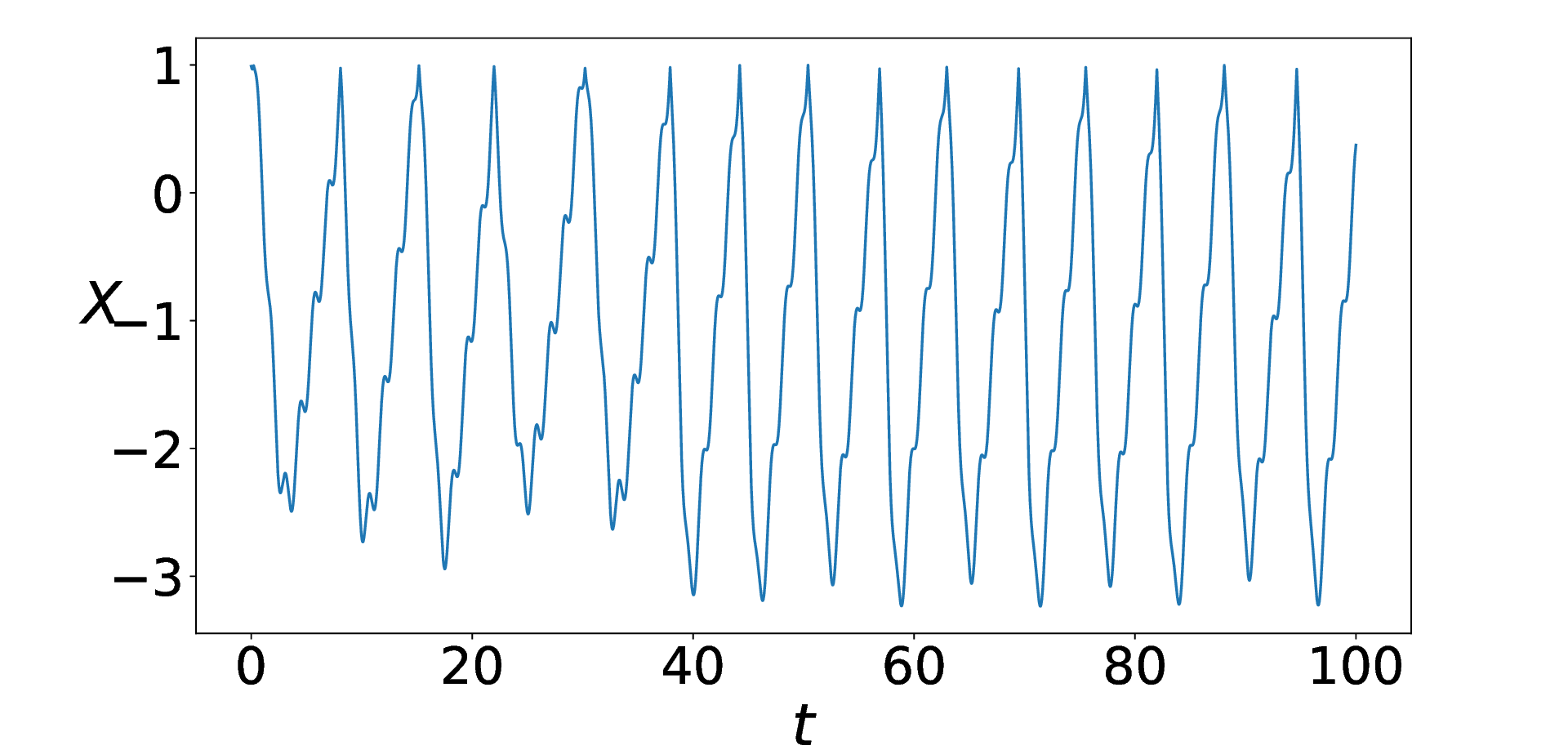}}
    \subcaptionbox{}{\includegraphics[scale=0.16]{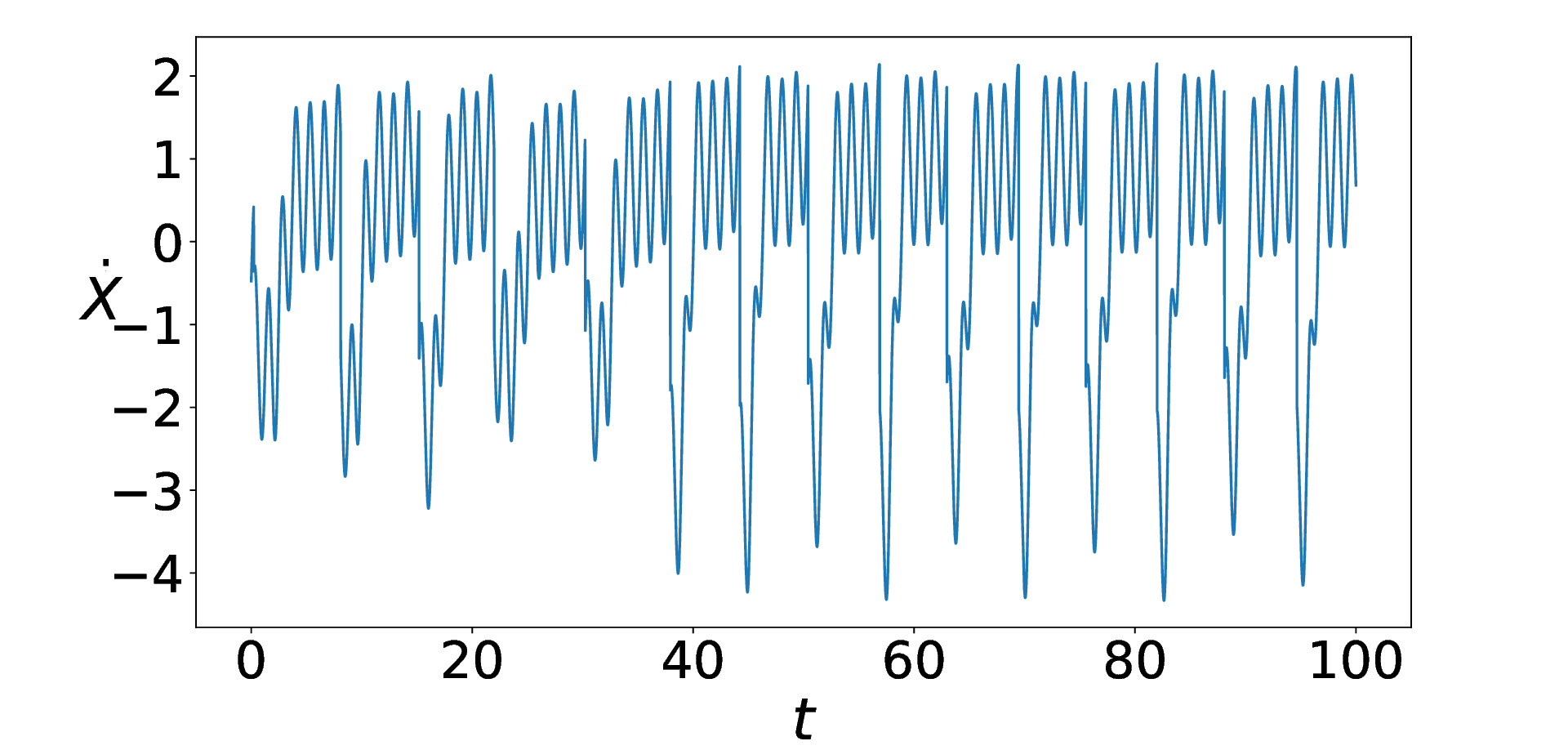}}
    \subcaptionbox{}{\includegraphics[scale=0.16]{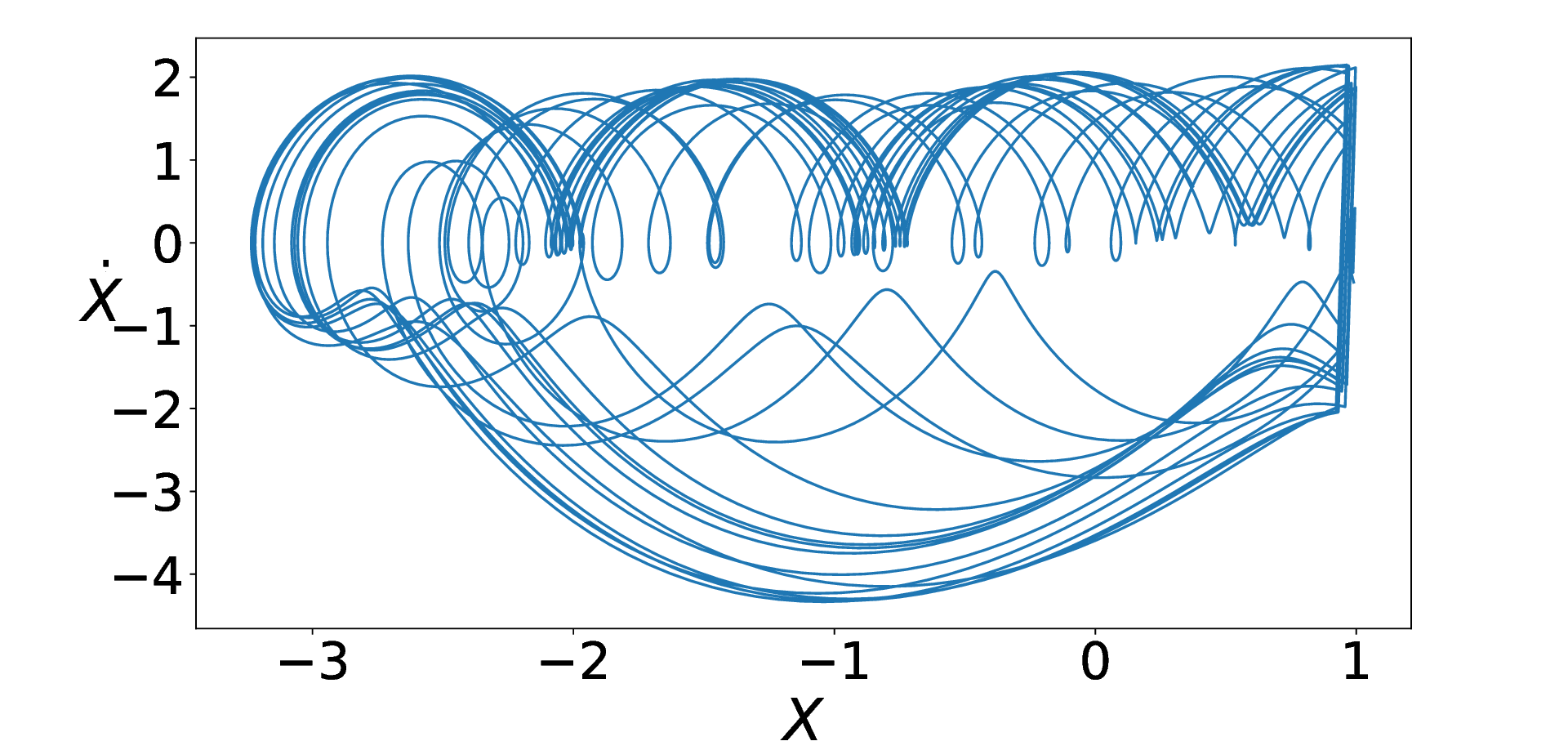}}

    \subcaptionbox{}{\includegraphics[scale=0.16]{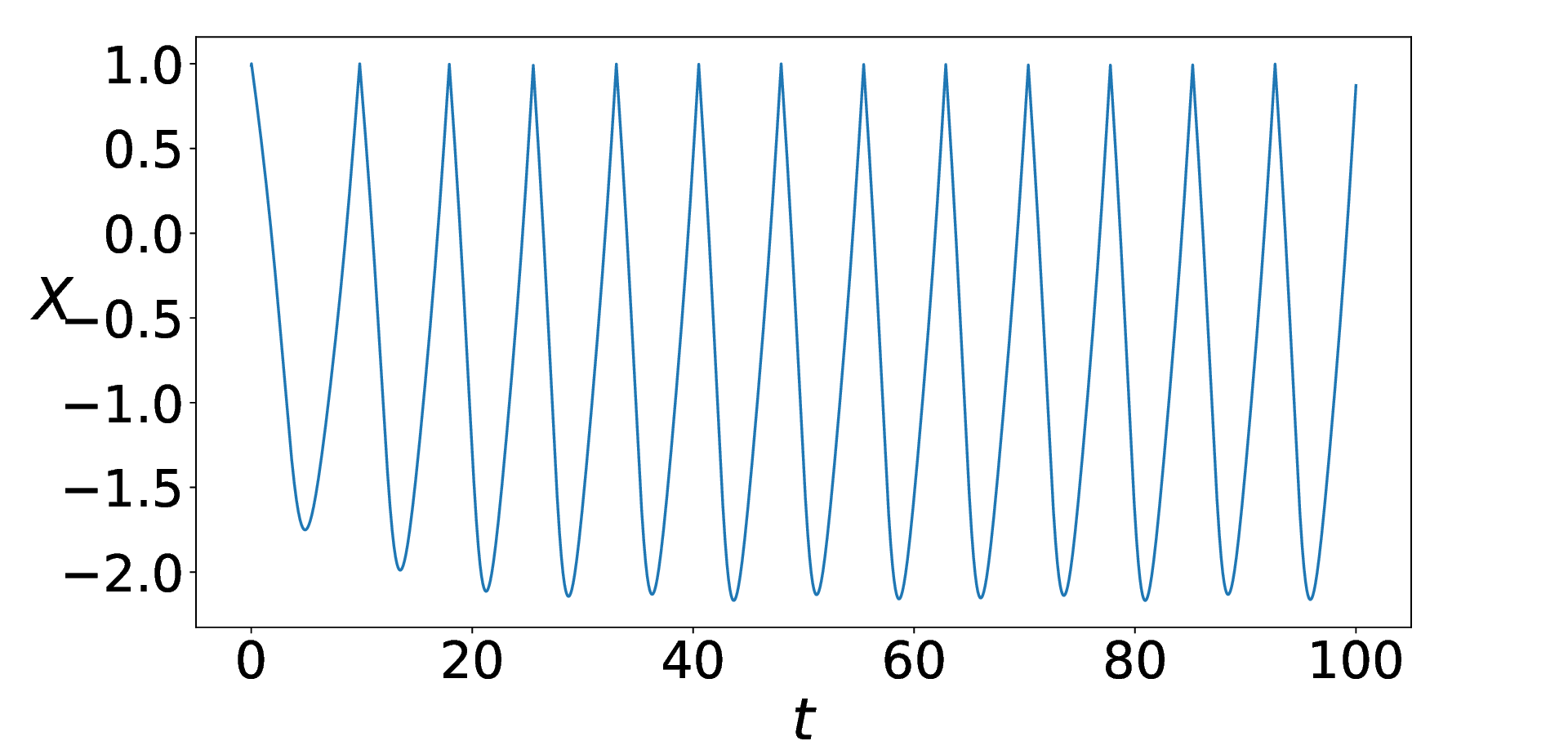}}
    \subcaptionbox{}{\includegraphics[scale=0.16]{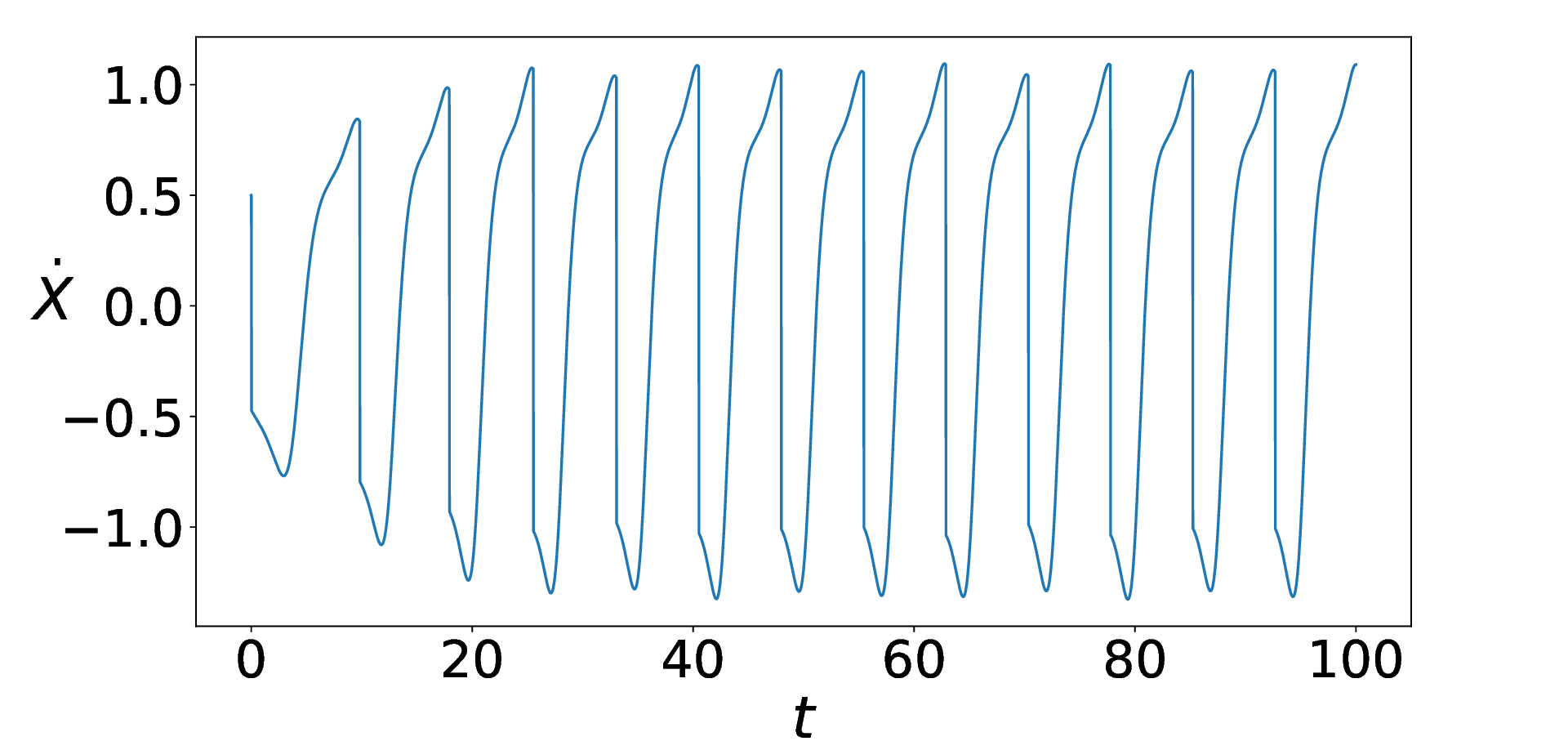}}
    \subcaptionbox{}{\includegraphics[scale=0.16]{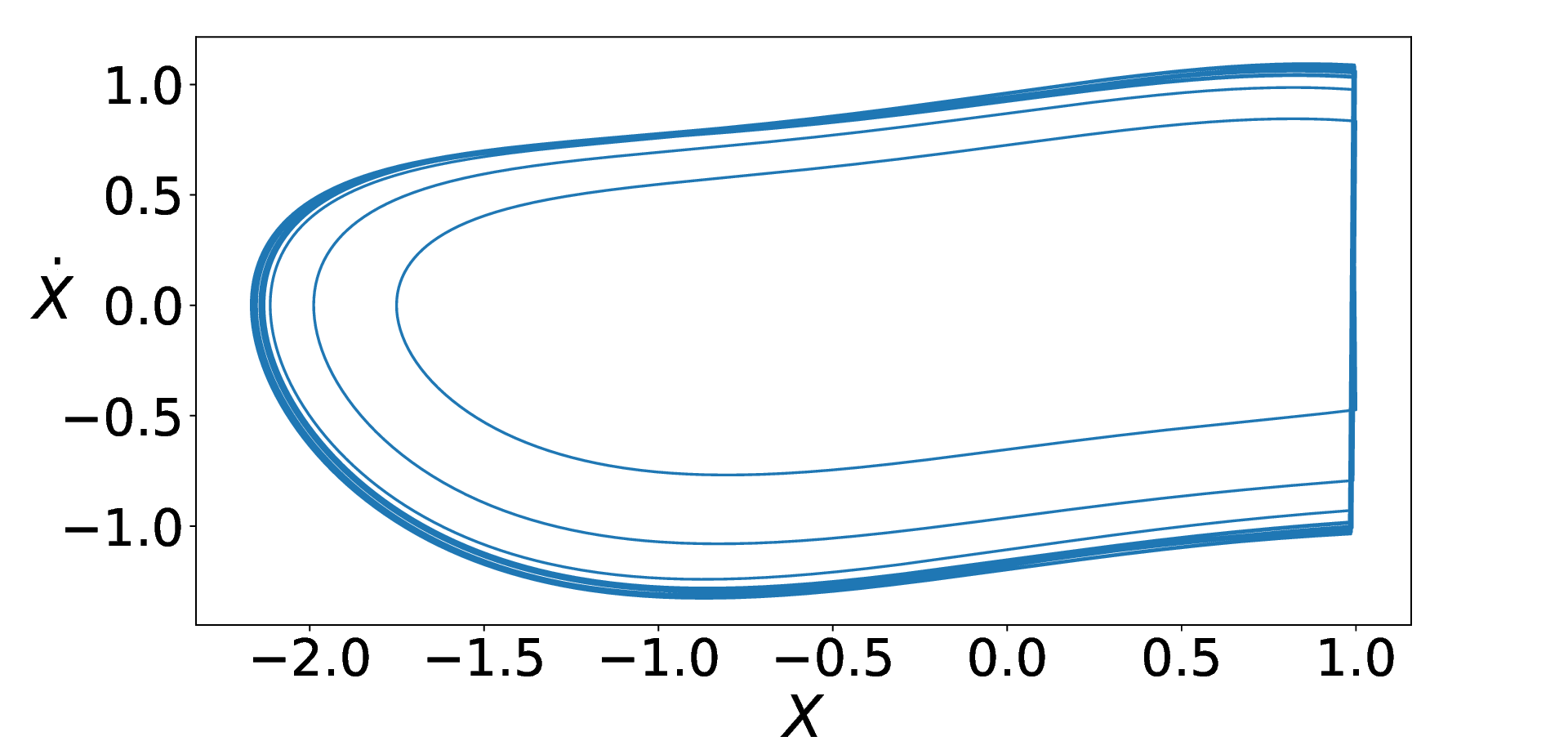}}
\caption{Comparison of original dynamics and effective dynamics for $g=5$ and $\Omega=5$. First row: simulations are based on the original system (Eqs.\eqref{eq2.1} and \eqref{eq2.2}). Second row: simulations are based on the effective system Eq.(\ref{eq2.10}). The pairs (a,d) represent position history, (b,e) depict velocity history, and (c,f) illustrate the phase plot of the corresponding original and effective slow dynamics.The other parameters are set at: $\gamma=0.22,\alpha=0.1,c=0.04,\omega_0=0.1,\omega=0.5, \Delta=1.0$ and $e=0.95$.The initial values of position and velocity are fixed at $x(0)=0.99$ and $\dot{x}(0)=0.5$.  }
\label{fig4}
\end{figure*}

\begin{figure*}
\centering
    \subcaptionbox{}{\includegraphics[height=2.8cm,width=5.3cm]{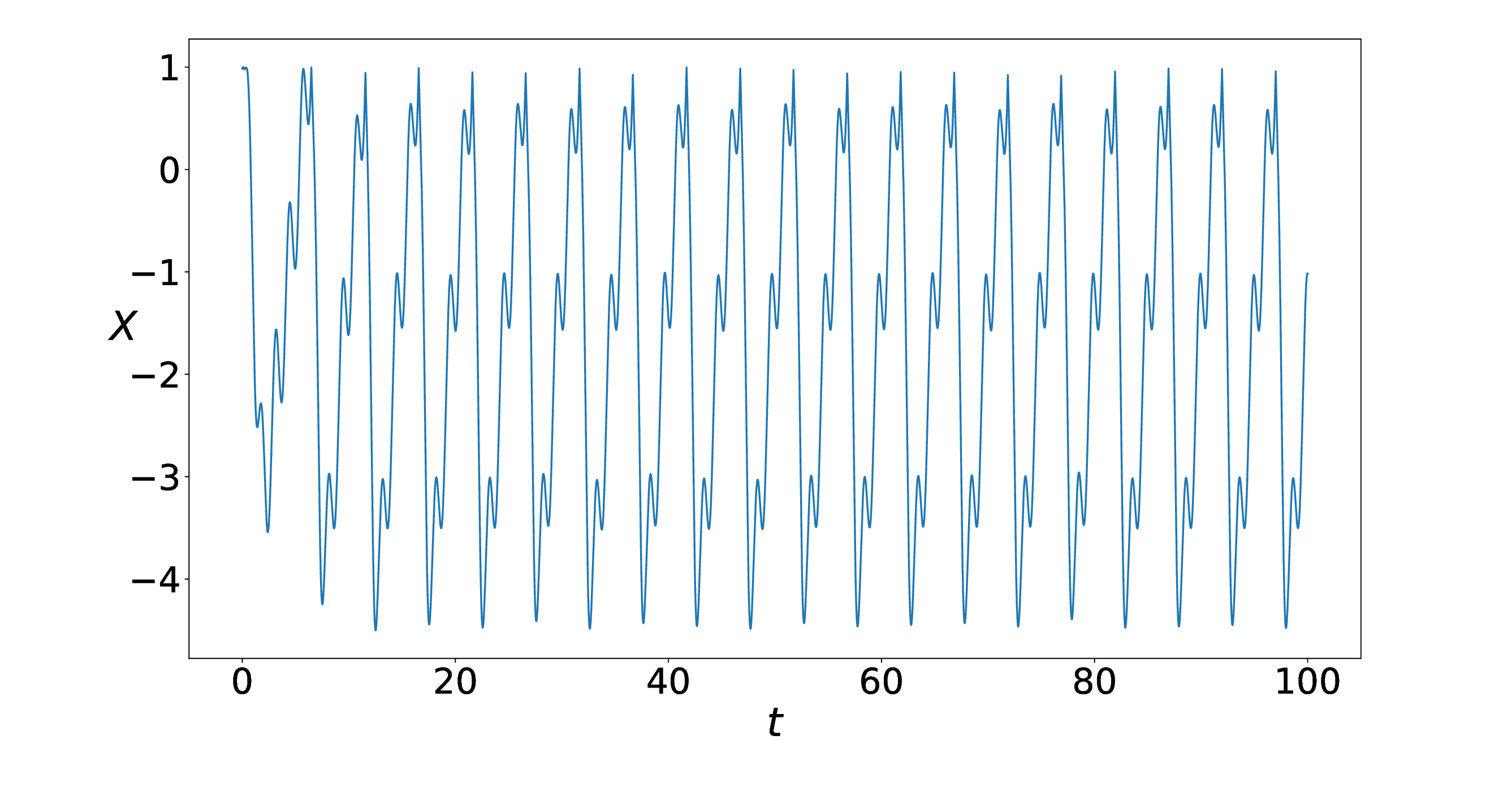}}
    \subcaptionbox{}{\includegraphics[height=2.8cm,width=5.3cm]{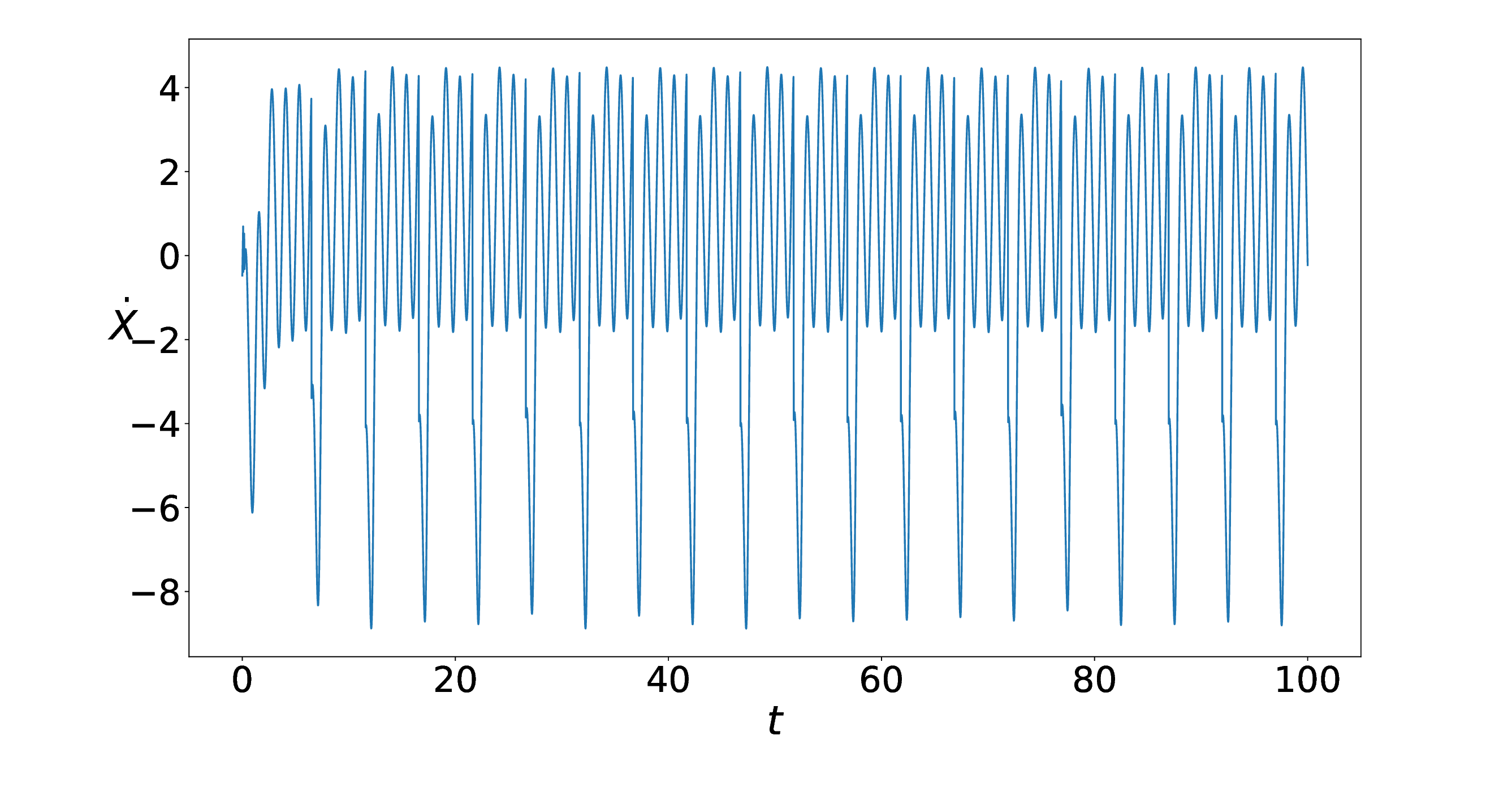}}
    \subcaptionbox{}{\includegraphics[height=2.8cm,width=5.3cm]{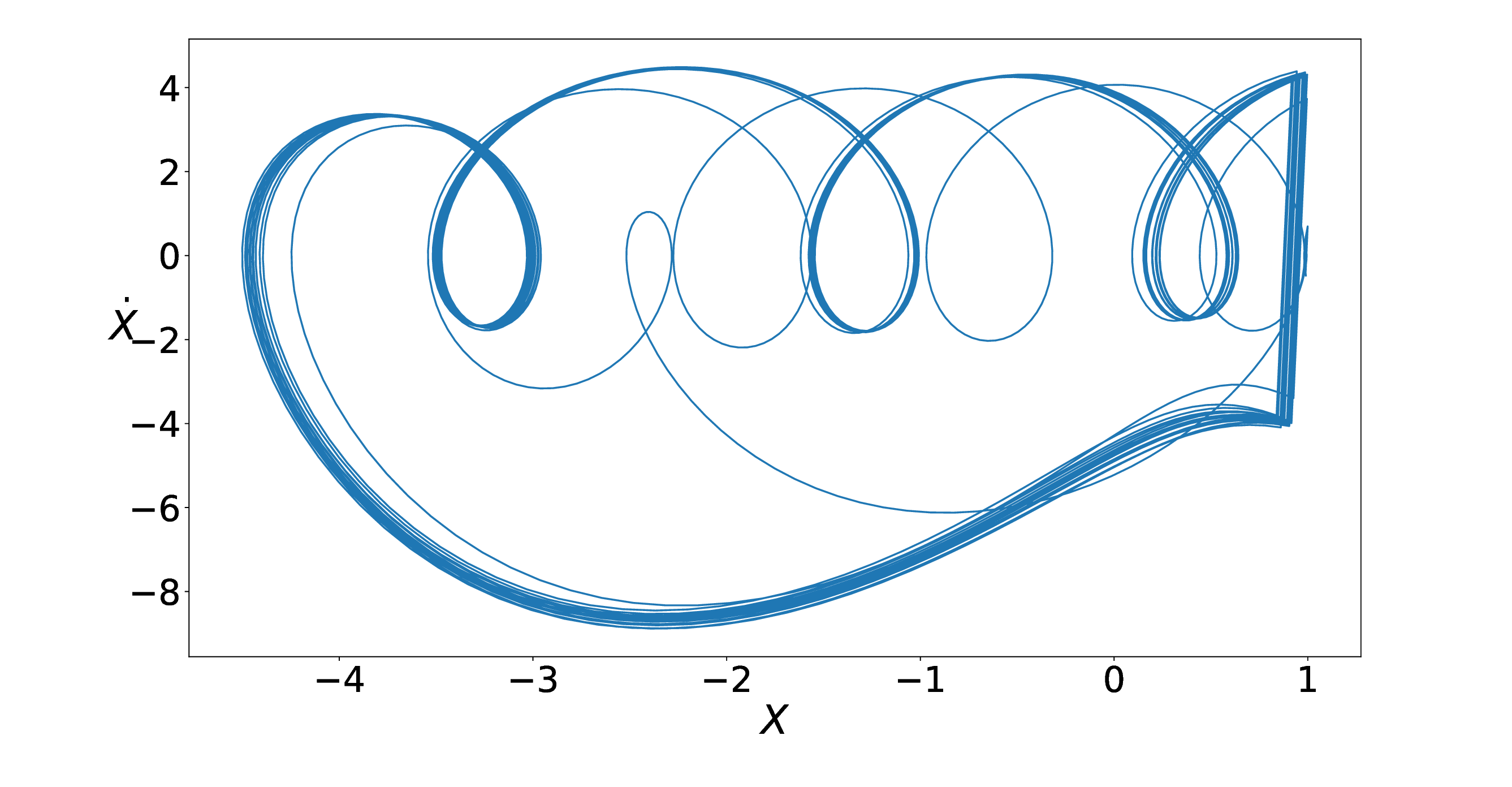}}

    \subcaptionbox{}{\includegraphics[height=2.8cm,width=5.3cm]{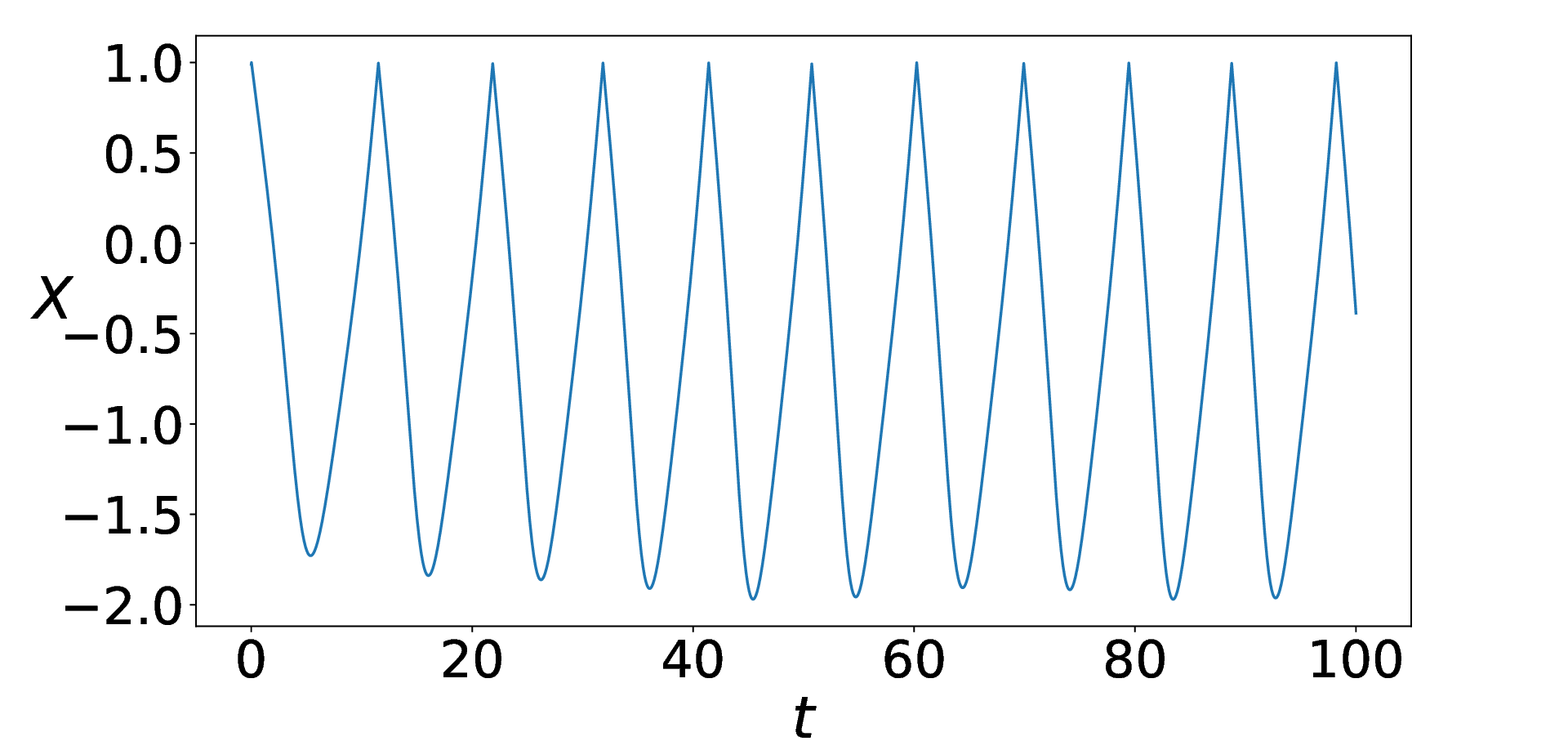}}
    \subcaptionbox{}{\includegraphics[height=2.8cm,width=5.3cm]{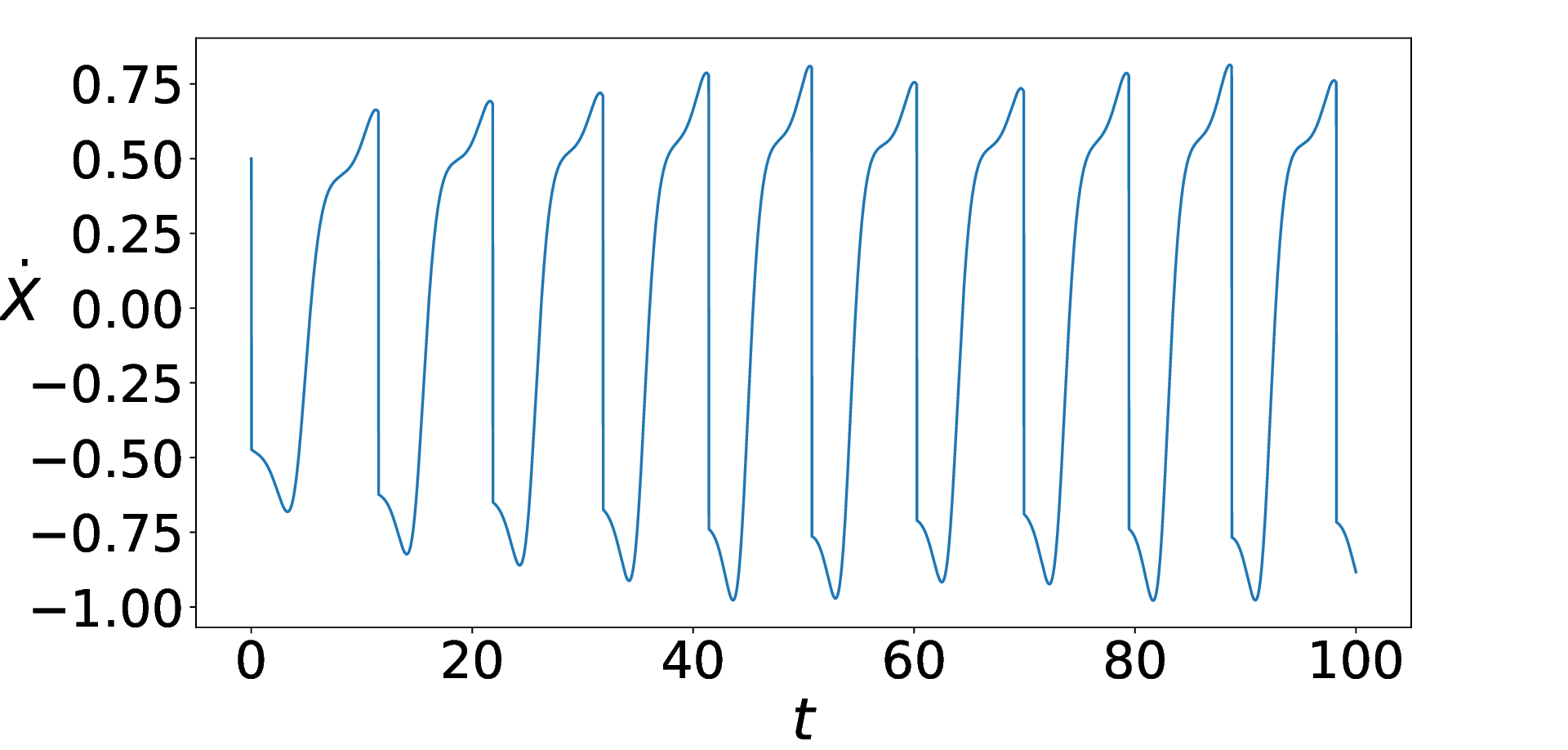}}
    \subcaptionbox{}{\includegraphics[height=2.8cm,width=5.3cm]{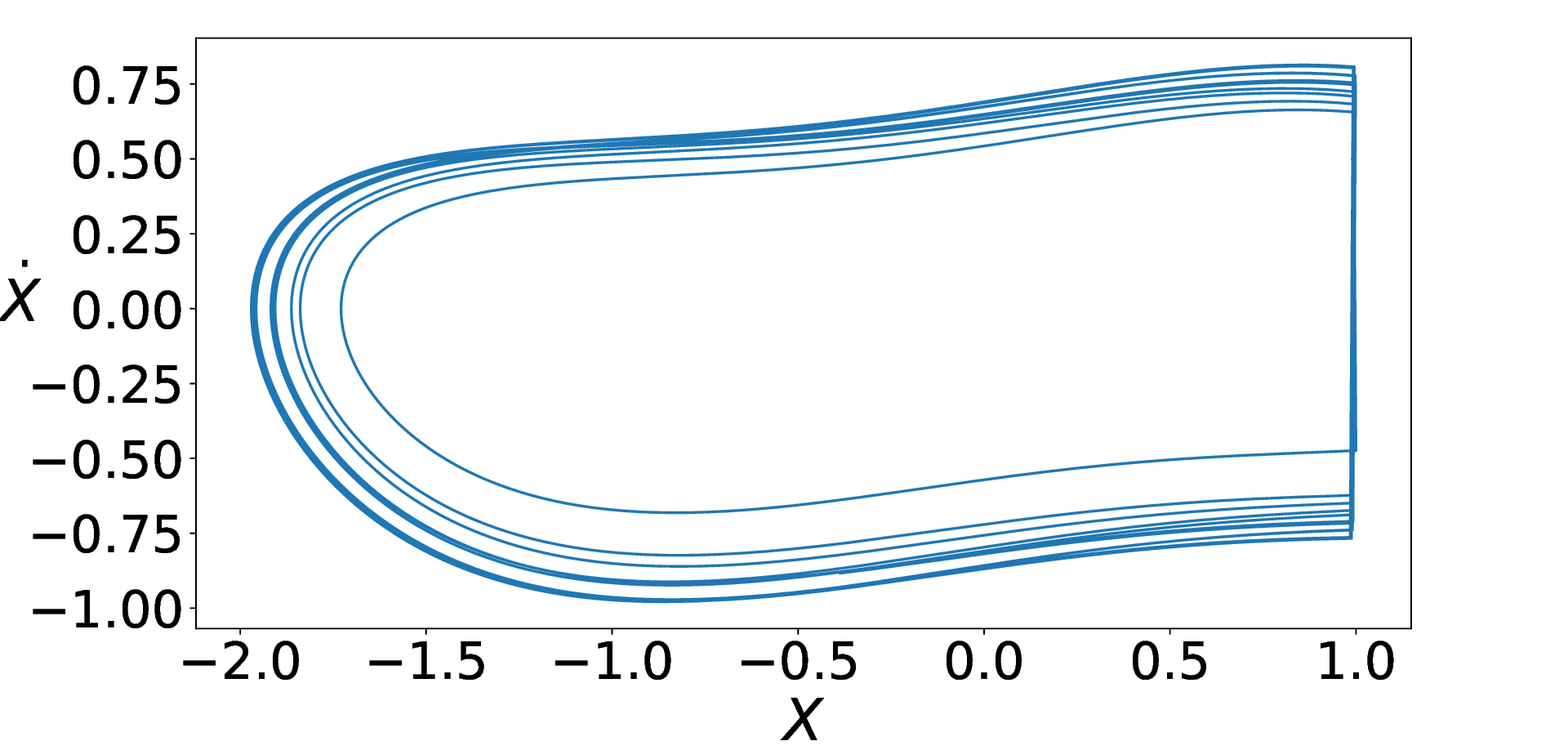}}
\caption{Comparison of dynamics at $g=15$ and $\Omega=5$.The top row shows simulations based on the original system (Eqs.\eqref{eq2.1} and \eqref{eq2.2}), while the bottom row depicts results from the effective system (Eq.\ref{eq2.10}). Each subplot illustrates: The pairs (a,d) represent position history, (b,e) depict velocity history, and (c,f) illustrate the phase plot of the corresponding original and effective slow dynamics.The other parameters include: $\gamma=0.22,\alpha=0.1,c=0.04,\omega_0=0.1,\omega=0.5, \Delta=1.0$ and $e=0.95$.The initial values of position and velocity are fixed at $x(0)=0.99$ and $\dot{x}(0)=0.5$.}
\label{fig5}
\end{figure*}

\begin{figure}
    \centering
    \subcaptionbox{}{\includegraphics[scale=0.49]{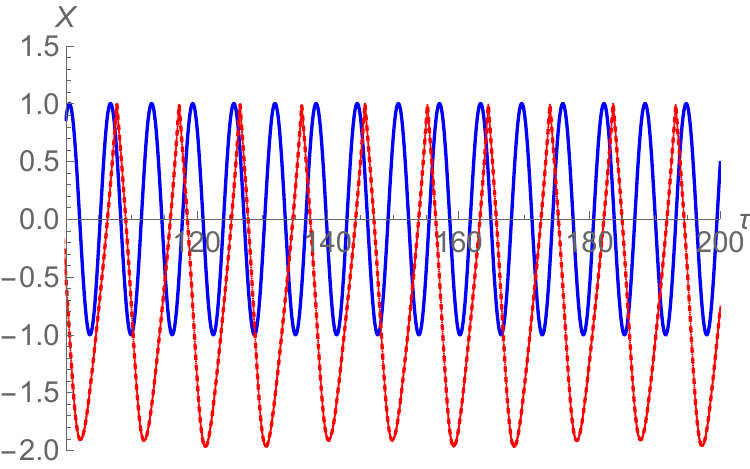}}
    \subcaptionbox{}{\includegraphics[scale=0.49]{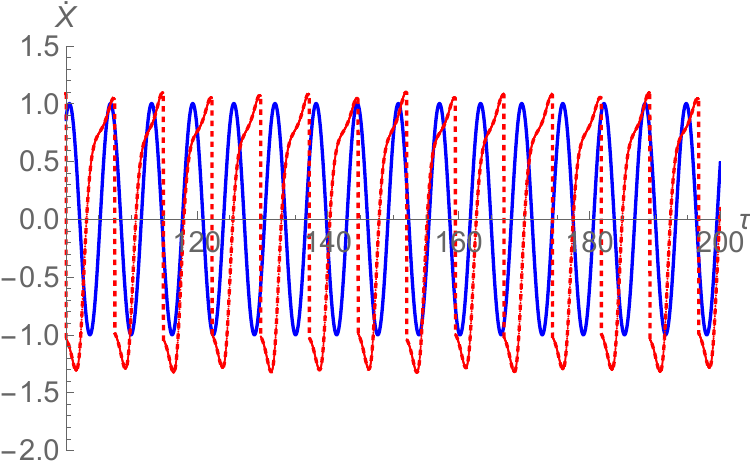}}
    \subcaptionbox{}{\includegraphics[scale=0.49]{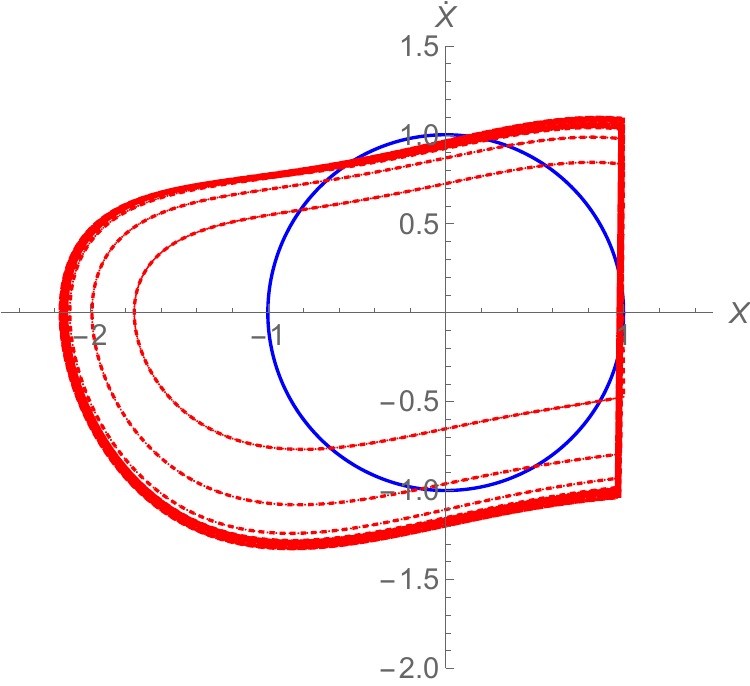}}
    \caption{Numerical (red dotted) and analytical (blue line) description of the (a) position-time, (b)  velocity-time and (c) position-velocity variation of the effective dynamics (Eq.(\ref{eq2.10})). Analytical result is obtained from Eq.(\ref{eq2.21}) for $g=5,\Omega=5,\Delta=1$.Other parameters are taken as:$\gamma=0.22,\alpha=0.1,c=0.04,\omega_0=0.1,\omega=0.5$ and $e=0.95$.The initial values of position and velocity are fixed at $x(0)=0.99$ and $\dot{x}(0)=0.5$}
    \label{fig6}
\end{figure}

Figs. \ref{fig3} and \ref{fig4} depict the time variation of position and velocity, as well as the phase portraits. The first row corresponds to the original dynamics simulated from Eqs. (\ref{eq2.1}) and (\ref{eq2.2}), and the second row corresponds to the effective slow dynamics. These plots are generated for a fixed value of $g=5$ and for two different values of $\Omega$, specifically $\Omega=4$ and $\Omega=5$, respectively.
Another noteworthy observation is the effect of increasing the fast frequency $\Omega$ from 4 to 5. The velocity of the original dynamics is reduced, and as the frequency increases, the original dynamics tend to settle down after an initial transient phase, exhibiting multi-periodicity in each case; see Figs.\ref{fig3}(c) and \ref{fig4}(c). However, when the fast forcing strength $g$ is increased to $15$ (see Fig.\ref{fig5}), the original system stabilizes almost immediately with a high rebound velocity (see Fig.\ref{fig5}(b)), though the periodicity decreases. In all these scenarios, the effective dynamics remain largely unchanged and continue to display periodic movements.

Notably, although the original dynamics exhibit multi-periodicity and multiple impacts, the effective dynamics clearly demonstrate sustained oscillations. The need for sustained oscillations should be emphasized due to several key factors. First, for resonance to occur, the system must accumulate energy from the driving excitation, enabling the amplitude of oscillation to build up over time. Sustained oscillations are also crucial for maintaining stable and predictable dynamics, ensuring that the system does not become chaotic and allowing for a clear observation of the resonance peak. Additionally, it helps to maintain phase locking with the driving force, ensuring continuous energy harvesting. Sustained oscillations of the effective dynamics thus facilitate energy accumulation, stability, and phase alignment, factors that are all essential for achieving and analyzing resonance in oscillatory systems. This can be visualized by the formation of a single-period limit cycle in the phase plots in the respective figures. This observation justifies the occurrence of resonance response, as the effective system shows sustained oscillatory behavior. Additionally, it is noticeable that as the fast frequency is increased, the size of the limit cycle contracts in the phase space. 

To investigate the effect of the increased strength of the fast forcing  on the system, another set of simulations have been carried out by tuning the amplitude up to the value $g=15$ and keeping the fast frequency fixed at $\Omega=5$, as illustrated in Fig. \ref{fig5}. The results indicate that the velocity of the original system attains higher values, reflecting a larger limit cycle region in the phase plot. Conversely, the effective dynamics remain in the sustained oscillation zone, displaying single periodicity. 
Fig.\ref{fig6} compares the dynamics derived from the slow motion (Eq.\eqref{eq2.10}) with the analytical predictions based on the flow equation in Eq.(\ref{eq2.24}) and the general solution of the system in Eq.(\ref{eq2.26}) using the two-time scale perturbation method. To determine $R(\tau_1)$ and $\theta(\tau_1)$,  Eq.(\ref{eq2.24}) is numerically integrated using the initial values $R_0$ and $\theta_0$. Substituting these solutions back into Eq.(\ref{eq2.21}) enables reconstructing up to the first order, the trajectory of the system, capturing the dominant term.
An inspection of these figures reveals the qualitative similarities between the analytical and numerical predictions. Moreover, notwithstanding the discrepancies in the quantitative values, the reasons for which have already been discussed, it is observed that both numerical and analytical methods converge on the same impact barrier position and also appear to have sustained oscillations. To improve accuracy and resolve frequency differences observed between the results, extending the perturbation analysis to higher orders would allow for further refinement by incorporating higher harmonics.

\section{Conclusion}
\noindent In this study, the resonant behavior of a vibro-impact oscillator with a one-sided barrier under simultaneous slow and fast forcing drives has been investigated. Detailed analytical techniques, such as Blekhman perturbation and multiple-scale analysis, alongside numerical simulations, are employed to examine how the strength of fast excitation $g$ influences the response of the oscillator to the weak forcing. It is demonstrated that a particular value of the fast forcing strength, $g_{res}$, maximizes the sensitivity of the system to the slow driving force 
$c\cos\omega t$, leading to the highest amplitude of oscillation. Control over $g$ provides a means to achieve increased system performance, improve experimental tractability, and provide an alternative approach to realize resonance in a nonsmooth oscillator. Future research could utilize more sophisticated analytical techniques, such as the Zhuravlev-Ivanov nonsmooth coordinate transformation, to better capture high-frequency dynamics and multi-impact scenarios. Additionally, advanced numerical methods could be employed to further understand the intricate dynamics at the boundary under high-frequency excitation. 

It is expected that these results can inspire new designs in the engineering realm, particularly in projects that benefit from precise control of resonance responses. By improving stimulus responses, these findings can be applied to mechanical devices such as energy harvesting technologies and microelectromechanical systems (MEMS) operating under the influence of weak excitation.

\section*{Acknowledgement}
SR acknowledges Dr. Aasifa Rounak for her valuable advice on the numerical processes. The authors are also thankful for the financial support received from the Ministry of Education, Govt of India towards IoE Projects Phase II for the project titled Complex Systems \& Dynamics.

-

\bibliographystyle{unsrt}

\begin{thebibliography}{10}

\bibitem{nayfeh2008applied}
Ali~H Nayfeh and Balakumar Balachandran.
\newblock {\em Applied nonlinear dynamics: analytical, computational, and experimental methods}.
\newblock John Wiley \& Sons, 2008.

\bibitem{manevich2005mechanics}
Arkadiy~I Manevich and Leonid~Isaakovich Manevich.
\newblock {\em The mechanics of nonlinear systems with internal resonances}.
\newblock World Scientific, 2005.

\bibitem{vyas2009microresonator}
Ashwin Vyas, Dimitrios Peroulis, and Anil~K Bajaj.
\newblock A microresonator design based on nonlinear 1: 2 internal resonance in flexural structural modes.
\newblock {\em Journal of Microelectromechanical Systems}, 18(3):744--762, 2009.

\bibitem{das2023nonlinear}
Rahul Das, Anil~K Bajaj, and Sayan Gupta.
\newblock Nonlinear energy sink coupled with a nonlinear oscillator.
\newblock {\em International Journal of Non-Linear Mechanics}, 148:104285, 2023.

\bibitem{das2024effectsinternalresonancedamping}
Rahul Das, Anil~K. Bajaj, and Sayan Gupta.
\newblock Effects of internal resonance and damping on koopman modes, 2024.

\bibitem{hajjaj2017mode}
Amal~Z Hajjaj, Md~Abdullah Hafiz, and Mohammad~I Younis.
\newblock Mode coupling and nonlinear resonances of mems arch resonators for bandpass filters.
\newblock {\em Scientific Reports}, 7(1):41820, 2017.

\bibitem{wang2019review}
Liang Wang, Weishan Chen, Junkao Liu, Jie Deng, and Yingxiang Liu.
\newblock A review of recent studies on non-resonant piezoelectric actuators.
\newblock {\em Mechanical Systems and Signal Processing}, 133:106254, 2019.

\bibitem{yang2024vibrational}
Jianhua Yang, S~Rajasekar, and Miguel~AF Sanju{\'a}n.
\newblock Vibrational resonance: A review.
\newblock {\em Physics Reports}, 1067:1--62, 2024.

\bibitem{landa2000vibrational}
PS~Landa and Peter~VE McClintock.
\newblock Vibrational resonance.
\newblock {\em Journal of Physics A: Mathematical and general}, 33(45):L433, 2000.

\bibitem{gitterman2001bistable}
M~Gitterman.
\newblock Bistable oscillator driven by two periodic fields.
\newblock {\em Journal of Physics A: Mathematical and General}, 34(24):L355, 2001.

\bibitem{blekhman2000vibrational}
Iliya~I Blekhman.
\newblock {\em Vibrational mechanics: nonlinear dynamic effects, general approach, applications}.
\newblock World Scientific, 2000.

\bibitem{blekhman2004conjugate}
II~Blekhman and PS~Landa.
\newblock Conjugate resonances and bifurcations in nonlinear systems under biharmonical excitation.
\newblock {\em International Journal of Non-Linear Mechanics}, 39(3):421--426, 2004.

\bibitem{baltanas2003experimental}
JP~Baltan{\'a}s, L~Lopez, II~Blechman, PS~Landa, A~Zaikin, J~Kurths, and MAF Sanju{\'a}n.
\newblock Experimental evidence, numerics, and theory of vibrational resonance in bistable systems.
\newblock {\em Physical Review E}, 67(6):066119, 2003.

\bibitem{roy2021vibrational}
Somnath Roy, Debapriya Das, and Dhruba Banerjee.
\newblock Vibrational resonance in a bistable van der pol--mathieu--duffing oscillator.
\newblock {\em International Journal of Non-Linear Mechanics}, 135:103771, 2021.

\bibitem{roy2023controlling}
Somnath Roy, Anirban Ray, and A~Roy Chowdhury.
\newblock Controlling subharmonic resonance and chaos by a fast forcing in a van der pol--duffing oscillator with parametrically excited damping.
\newblock {\em Chaos, Solitons \& Fractals}, 174:113857, 2023.

\bibitem{jeevarathinam2013effect}
C~Jeevarathinam, S~Rajasekar, and MAF Sanju{\'a}n.
\newblock Effect of multiple time-delay on vibrational resonance.
\newblock {\em Chaos: An Interdisciplinary Journal of Nonlinear Science}, 23(1), 2013.

\bibitem{chowdhury2020weak}
Avishek Chowdhury, Marcel~G Clerc, Sylvain Barbay, Isabelle Robert-Philip, and Remy Braive.
\newblock Weak signal enhancement by nonlinear resonance control in a forced nano-electromechanical resonator.
\newblock {\em Nature Communications}, 11(1):2400, 2020.

\bibitem{deng2010vibrational}
Bin Deng, Jiang Wang, Xile Wei, KM~Tsang, and Wai~Lok Chan.
\newblock Vibrational resonance in neuron populations.
\newblock {\em Chaos: An Interdisciplinary Journal of Nonlinear Science}, 20(1), 2010.

\bibitem{ibrahim2009vibro}
Raouf~A Ibrahim.
\newblock {\em Vibro-Impact Dynamics: Modeling, Mapping and Applications}, volume~43.
\newblock Springer Science \& Business Media, 2009.

\bibitem{dimentberg2004random}
MF~Dimentberg and DV~Iourtchenko.
\newblock Random vibrations with impacts: a review.
\newblock {\em Nonlinear Dynamics}, 36:229--254, 2004.

\bibitem{gu2018dynamical}
XD~Gu and Z~CH Deng.
\newblock Dynamical analysis of vibro-impact capsule system with hertzian contact model and random perturbation excitations.
\newblock {\em Nonlinear Dynamics}, 92:1781--1789, 2018.

\bibitem{de2019drill}
Luciano~PP de~Moraes and Marcelo~A Savi.
\newblock Drill-string vibration analysis considering an axial-torsional-lateral nonsmooth model.
\newblock {\em Journal of Sound and Vibration}, 438:220--237, 2019.

\bibitem{jangid2001base}
RS~Jangid and JM~Kelly.
\newblock Base isolation for near-fault motions.
\newblock {\em Earthquake Engineering \& Structural Dynamics}, 30(5):691--707, 2001.

\bibitem{dimitrakopoulos2013nonsmooth}
Elias~G Dimitrakopoulos.
\newblock Nonsmooth analysis of the impact between successive skew bridge-segments.
\newblock {\em Nonlinear Dynamics}, 74:911--928, 2013.

\bibitem{kumar2015calculation}
ANDREW~S. WHITTAKER, MANISH KUMAR, and MANISH KUMAR.
\newblock Seismic isolation of nuclear power plants.
\newblock {\em Nuclear Engineering and Technology}, 46(5):569--580, 2014.

\bibitem{luo2008periodic}
Guanwei Luo, Jianhua Xie, Xifeng Zhu, and Jiangang Zhang.
\newblock Periodic motions and bifurcations of a vibro-impact system.
\newblock {\em Chaos, Solitons \& Fractals}, 36(5):1340--1347, 2008.

\bibitem{zhang2017detecting}
Yongxiang Zhang and Guanwei Luo.
\newblock Detecting unstable periodic orbits and unstable quasiperiodic orbits in vibro-impact systems.
\newblock {\em International Journal of Non-Linear Mechanics}, 96:12--21, 2017.

\bibitem{blazejczyk2001analysis}
Barbara Blazejczyk-Okolewska.
\newblock Analysis of an impact damper of vibrations.
\newblock {\em Chaos, Solitons \& Fractals}, 12(11):1983--1988, 2001.

\bibitem{luo1998hopf}
G-W Luo and J-H Xie.
\newblock Hopf bifurcation of a two-degree-of-freedom vibro-impact system.
\newblock {\em Journal of Sound and Vibration}, 213(3):391--408, 1998.

\bibitem{luo2002hopf}
GW~Luo and JH~Xie.
\newblock Hopf bifurcations and chaos of a two-degree-of-freedom vibro-impact system in two strong resonance cases.
\newblock {\em International Journal of Non-Linear Mechanics}, 37(1):19--34, 2002.

\bibitem{yue2008symmetry}
Y~Yue and JH~Xie.
\newblock Symmetry and bifurcations of a two-degree-of-freedom vibro-impact system.
\newblock {\em Journal of Sound and Vibration}, 314(1-2):228--245, 2008.

\bibitem{de2008control}
Silvio~LT De~Souza, Iber{\^e}~L Caldas, Ricardo~L Viana, and Jos{\'e}~M Balthazar.
\newblock Control and chaos for vibro-impact and non-ideal oscillators.
\newblock {\em Journal of Theoretical and Applied Aechanics}, 46(3):641--664, 2008.

\bibitem{luo2001bifurcations}
Guanwei Luo and Jianhua Xie.
\newblock Bifurcations and chaos in a system with impacts.
\newblock {\em Physica D: Nonlinear Phenomena}, 148(3-4):183--200, 2001.

\bibitem{feng2015chaotic}
Jinqian Feng and Junli Liu.
\newblock Chaotic dynamics of the vibro-impact system under bounded noise perturbation.
\newblock {\em Chaos, Solitons \& Fractals}, 73:10--16, 2015.

\bibitem{yang2017stochastic}
Yongge Yang, Wei Xu, Yahui Sun, and Yanwen Xiao.
\newblock Stochastic bifurcations in the nonlinear vibroimpact system with fractional derivative under random excitation.
\newblock {\em Communications in Nonlinear Science and Numerical Simulation}, 42:62--72, 2017.

\bibitem{kumar2016stochastic}
Pankaj Kumar, S~Narayanan, and Sayan Gupta.
\newblock Stochastic bifurcations in a vibro-impact duffing--van der pol oscillator.
\newblock {\em Nonlinear Dynamics}, 85:439--452, 2016.

\bibitem{kumar2017bifurcation}
Pankaj Kumar, S~Narayanan, and Sayan Gupta.
\newblock Bifurcation analysis of a stochastically excited vibro-impact duffing-van der pol oscillator with bilateral rigid barriers.
\newblock {\em International Journal of Mechanical Sciences}, 127:103--117, 2017.

\bibitem{qian2021stochastic}
Jiamin Qian and Lincong Chen.
\newblock Stochastic p-bifurcation analysis of a novel type of unilateral vibro-impact vibration system.
\newblock {\em Chaos, Solitons \& Fractals}, 149:111112, 2021.

\bibitem{rounak2020stochastic}
Aasifa Rounak and Sayan Gupta.
\newblock Stochastic p-bifurcation in a nonlinear impact oscillator with soft barrier under ornstein--uhlenbeck process.
\newblock {\em Nonlinear Dynamics}, 99(4):2657--2674, 2020.

\bibitem{makarenkov2012dynamics}
Oleg Makarenkov and Jeroen~SW Lamb.
\newblock Dynamics and bifurcations of nonsmooth systems: A survey.
\newblock {\em Physica D: Nonlinear Phenomena}, 241(22):1826--1844, 2012.

\bibitem{belykh2023beyond}
Igor Belykh, Rachel Kuske, Maurizio Porfiri, and David~JW Simpson.
\newblock Beyond the bristol book: Advances and perspectives in non-smooth dynamics and applications.
\newblock {\em Chaos: An Interdisciplinary Journal of Nonlinear Science}, 33(1), 2023.

\bibitem{luo2007vibro}
Guanwei Luo, Yanlong Zhang, Jianhua Xie, and Jiangang Zhang.
\newblock Vibro-impact dynamics near a strong resonance point.
\newblock {\em Acta Mechanica Sinica}, 23(3):329--341, 2007.

\bibitem{rong2010resonant}
Haiwu Rong, Xiangdong Wang, Wei Xu, and Tong Fang.
\newblock Resonant response of a non-linear vibro-impact system to combined deterministic harmonic and random excitations.
\newblock {\em International Journal of Non-Linear Mechanics}, 45(5):474--481, 2010.

\bibitem{kember1999excitation}
SA~Kember and VI~Babitsky.
\newblock Excitation of vibro-impact systems by periodic impulses.
\newblock {\em Journal of sound and vibration}, 227(2):427--447, 1999.

\bibitem{fritzkowski2021near}
Pawel Fritzkowski and Jan Awrejcewicz.
\newblock Near-resonant dynamics, period doubling and chaos of a 3-dof vibro-impact system.
\newblock {\em Nonlinear Dynamics}, 106(1):81--103, 2021.

\bibitem{cho2013vibro}
Tongjun Cho.
\newblock Vibro-acoustic characteristics of floating floor system: The influence of frequency-matched resonance on low frequency impact sound.
\newblock {\em Journal of Sound and Vibration}, 332(1):33--42, 2013.

\bibitem{luo2008dynamics}
GW~Luo, XH~Lv, and XF~Zhu.
\newblock Dynamics of vibro-impact mechanical systems with large dissipation.
\newblock {\em International Journal of Mechanical Sciences}, 50(2):214--232, 2008.

\bibitem{nguyen2017new}
Van-Du Nguyen, Huu-Cong Nguyen, Nhu-Khoa Ngo, and Ngoc-Tuan La.
\newblock A new design of horizontal electro-vibro-impact devices.
\newblock {\em Journal of Computational and Nonlinear Dynamics}, 12(6):061002, 2017.

\bibitem{haroun2015micro}
Ahmed Haroun, Ichiro Yamada, and Shin’ichi Warisawa.
\newblock Micro electromagnetic vibration energy harvester based on free/impact motion for low frequency--large amplitude operation.
\newblock {\em Sensors and Actuators A: Physical}, 224:87--98, 2015.

\bibitem{haiwu2009subharmonic}
Rong Haiwu, Xiangdong Wang, Wei Xu, and Tong Fang.
\newblock Subharmonic response of a single-degree-of-freedom nonlinear vibro-impact system to a narrow-band random excitation.
\newblock {\em Physical Review E—Statistical, Nonlinear, and Soft Matter Physics}, 80(2):026604, 2009.

\bibitem{superharmonic}
Amine Bichri and Mohamed Belhaq.
\newblock {Control of a Forced Impacting Hertzian Contact Oscillator Near Sub- and Superharmonic Resonances of Order 2}.
\newblock {\em Journal of Computational and Nonlinear Dynamics}, 7(1):011003, 07 2011.

\bibitem{tsai2022cmos}
Chun-Pu Tsai and Wei-Chang Li.
\newblock Cmos-mems vibro-impact devices and applications.
\newblock {\em Frontiers in Mechanical Engineering}, 8:898328, 2022.

\bibitem{han2017study}
Yanhui Han, Yue Feng, Zejie Yu, Wenzhong Lou, and Huicong Liu.
\newblock A study on piezoelectric energy-harvesting wireless sensor networks deployed in a weak vibration environment.
\newblock {\em IEEE Sensors Journal}, 17(20):6770--6777, 2017.

\bibitem{liu2023harvesting}
Qi~Liu, Weiyang Qin, Yongfeng Yang, and Zhiyong Zhou.
\newblock Harvesting weak vibration energy by amplified inertial force and multi-stable buckling piezoelectric structure.
\newblock {\em Mechanical Systems and Signal Processing}, 189:110125, 2023.

\bibitem{shi2015study}
Huichao Shi, Shangchun Fan, Weiwei Xing, and Jinhao Sun.
\newblock Study of weak vibrating signal detection based on chaotic oscillator in mems resonant beam sensor.
\newblock {\em Mechanical systems and signal processing}, 50:535--547, 2015.

\bibitem{bo2014weak}
Liu Bo, Wang~You Bo, and Fan JianYing.
\newblock Weak signal acquisition and processing system for piezoelectric ceramic transducer actuator.
\newblock In {\em 2014 9th International Forum on Strategic Technology (IFOST)}, pages 209--215. IEEE, 2014.

\bibitem{venkatesan1997bifurcation}
A~Venkatesan and M~Lakshmanan.
\newblock Bifurcation and chaos in the double-well duffing--van der pol oscillator: numerical and analytical studies.
\newblock {\em Physical Review E}, 56(6):6321, 1997.

\bibitem{xu2020independent}
Yeyin Xu and Albert~CJ Luo.
\newblock Independent period-2 motions to chaos in a van der pol--duffing oscillator.
\newblock {\em International Journal of Bifurcation and Chaos}, 30(15):2030045, 2020.

\bibitem{wang2009response}
Liang Wang, Wei Xu, Gaojie Li, and Dongxi Li.
\newblock Response of a stochastic duffing--van der pol elastic impact oscillator.
\newblock {\em Chaos, Solitons \& Fractals}, 41(4):2075--2080, 2009.

\bibitem{kumar2017modified}
Prakash Kumar, Anil Kumar, and Silvano Erlicher.
\newblock A modified hybrid van der pol--duffing--rayleigh oscillator for modelling the lateral walking force on a rigid floor.
\newblock {\em Physica D: Nonlinear Phenomena}, 358:1--14, 2017.

\bibitem{chakraborty2021nonlinear}
G~Chakraborty and Nikul Jani.
\newblock Nonlinear dynamics of resonant microelectromechanical system (mems): A review.
\newblock {\em Mechanical Sciences: The Way Forward}, pages 57--81, 2021.

\end{thebibliography}

\end{document}